\documentclass[12 pt, a4paper]{article}

\usepackage{graphicx}
\usepackage[utf8]{inputenc}
\usepackage[T1]{fontenc}
\usepackage{lmodern}
\usepackage{geometry}
\usepackage{amsmath} 
\allowdisplaybreaks
\usepackage{amsthm}
\usepackage{amssymb}
\usepackage{float}
\usepackage{color}
\usepackage[english]{babel}
\usepackage[ruled,vlined]{algorithm2e}
\usepackage{multirow}
\usepackage{csquotes}
\usepackage{setspace}
\usepackage{longtable}
\usepackage{caption}
\usepackage{subcaption}
\usepackage{tikz}
\usepackage{emptypage}
\usepackage{bm}
\usepackage{hhline}
\usepackage{diagbox}
\usepackage[colorlinks, citecolor=blue]{hyperref}
\usepackage[authoryear, round]{natbib}
\usepackage{authblk}
\usepackage{booktabs}
\usepackage{hyperref}

\title{Evaluating multi-season occupancy models with autocorrelation fitted to heterogeneous datasets}
\author[1,2]{Andr\'e Lu\'is Luza}
\author[1,3]{Didier Alard}
\author[2]{Fr\'ed\'eric Barraquand}
\affil[1]{UMR Biodiversité Gènes et Communautés, University of Bordeaux,  INRAE, Pessac, France}
\affil[2]{Institute of Mathematics of Bordeaux, University of Bordeaux, CNRS, Bordeaux INP, Talence, France}
\affil[3]{US Fauna, University of Bordeaux, Pessac, France}
\date{}

\begin{document}

\maketitle


\begin{abstract}

    Predicting species distributions using occupancy models accounting for imperfect detection is now commonplace in ecology. Recently, modeling spatial and temporal autocorrelation was proposed to alleviate the lack of replication in occupancy data, which often prevents model identifiability. However, how such models perform in highly heterogeneous datasets where missing or single-visit data dominates remains an open question. Motivated by a heterogeneous fine-scale butterfly occupancy dataset, we evaluate the performance of a multi-season occupancy model with spatial and temporal random effects to a skewed (Poisson) distribution of the number of surveys per site, overlap of covariates between occupancy and detection submodels, and spatiotemporal clustering of observations. Results showed that the model is robust to heterogeneous data and covariate overlap. However, when spatiotemporal gaps were added, site occupancy was biased towards the average occupancy, itself overestimated. Random effects did not correct the influence of gaps, due to identifiability issues of variance and autocorrelation parameters. Occupancy analysis of two butterfly species further confirmed these results. Overall, multi-season occupancy models with autocorrelation are robust to heterogeneous data and covariate overlap, but still present identifiability issues and are challenged by severe data gaps, which contaminate predictions even in data-rich areas.
    
\end{abstract}

\textbf{Keywords}: opportunistic data, occupancy, species distribution models, spatial random effects, temporal random effects, identifiability\\

\noindent  {\small  $^*$ Correspondence to \texttt{luza.andre@gmail.com}, \texttt{frederic.barraquand@u-bordeaux.fr}}

\thispagestyle{empty}

\newpage
\doublespacing

\section{Introduction}

Heterogeneous databases, where various data types are pooled together, have increasingly been used to predict species distributions and trends with site occupancy models \citep{hochachka2023considerations, vonhirschheydt2023mixed}. Occupancy models allow in theory to produce not only maps of presence but also maps of probability of detection \citep{kery2013analysing}, which can minimize estimation and prediction bias, identify the sources of distributional uncertainties, and guide future data collection \citep{lahoz-monfort2014imperfect, guillera-arroita2017modelling}.

However, in large heterogeneous naturalist databases, a lot of the data are single-visit data or even missing data, i.e., non-visited cells \citep{kelling2019semistructured, johnston2020estimating}. This poses a challenge for occupancy models, that are typically identifiable (i.e., a unique set of parameters can be estimated from the data, \citealp{cole20202parameter}) when fitted to data following the robust sampling design. Under the robust sampling design, repeated visits are carried out over multiple secondary occasions within a given primary occasion (during which occupancy is assumed to be constant) to all or some sites, thus allowing for separate estimation of occupancy and detection \citep{mackenzie2002estimating, mackenzie2003estimating, mackenzie2005designing, guillera-arroita2010design, knape2015estimates, reich2020optimal}. Nonetheless, the broad availability of opportunistic observations (snapshots in space and time) motivated the use of abundance and occupancy models to single-survey data. Simulations and case studies show that, given a large number of sampled sites and years, and no overlap of covariates influencing the actual occupancy $\psi$ and the detection probability $p$, the occupancy model might be identifiable and estimable \citep{lele2012dealing, solymos2016revisiting, peach2017single}. However, the identifiability of this model is fragile and sampling design recommendations usually include repeated secondary occasions \citep{mackenzie2005designing, guillera-arroita2010design, knape2015estimates, reich2020optimal}. The approach has also been criticized because assumptions about non-overlapping covariates are hardly met in practice, since often the same covariate might affect both occupancy and detection (\citealp{ ruiz-gutierrez2010occupancy, lahoz-monfort2014imperfect}).

Recently, research has leveraged the possibility of using spatial and temporal random effects to represent spatial and temporal autocorrelation in multi-season site-occupancy models \citep{hepler2018identifying,hepler2021spatiotemporal,diana2023fast,doser2024fractional}. In these models, the autocorrelations will transmit the information that a focal site will resemble neighboring sites in space and time, and therefore share similar values of occupancy or detection probabilities. This property has been dubbed ``fractional replication'' by \citet{doser2024fractional}. This use of autocorrelation as a substitute for a strict adherence to the robust design, with repeats within primary occasions, offers interesting avenues to analyze large heterogeneous occupancy datasets comprising a skewed distribution of the number of visits per grid cell. \citet{doser2024fractional}, hereafter D\&S, showed using simulations that their ``fractional replication'' model is identifiable with single-visit data under strict parametric assumptions, as well as robustly so (no dependence on exact parametric assumptions and the correct model specification) when a small fraction (10\%) of repeated visits is added to sites within primary occasions. 

Despite these fruitful developments, methods using fractional replication remain in their infancy. Based on our exploration of a large public occupancy dataset of butterfly species in the French Southwest, we identify a number of challenges to the methods that require extending the simulation study of D\&S. First, in this dataset and likely most fine-scale occupancy data, the number of visits per cell exhibits a Poisson-like distribution starting at 0 (including grid cells with NAs) rather than a two-group mixture of cells visited once and cells revisited a fixed number of times, as used in D\&S and recent evaluations of occupancy models such as \citet{vonhirschheydt2023mixed}. Second, the covariates affecting occupancy and detection in D\&S were fully random (uncorrelated) in space and time, as well as with regard to each other, and were designed to vary in both space and time. This puts the model in a very optimistic scenario. In many real datasets, the covariates will be spatially autocorrelated, some will jointly affect detection and occupancy probabilities, and some will vary along a single dimension \citep[either space or time,][]{ruiz-gutierrez2010occupancy}, constraints already shown to be a challenge for occupancy models \citep{royle2006site, lele2012dealing, peach2017single}. Third, in many datasets (such as those of butterflies) phenology will group observations at specific times \citep{matechou2014monitoring, strebel2014studying} and observers' behavior will group observations at specific places and times \citep{altwegg2019occupancy, johnston2020estimating}, further complexifying the inference. Here, we progressively incorporate those ecologically-motivated constraints into the performance assessment of a multi-season occupancy model with spatial and temporal autocorrelation fitted to heterogeneous datasets. 

\section{Model and methods}

\subsection{Motivating empirical example}

We modeled species distribution using a compilation of data from 504 projects (data sources) of standardized and opportunistic butterfly records obtained in the Nouvelle-Aquitaine region, Southwest France, from 2000 to 2023. These data have been compiled by the Nouvelle-Aquitaine Wildlife Observatory (Observatoire de la faune sauvage de Nouvelle-Aquitaine - FAUNA, https://observatoire-fauna.fr/; Université de Bordeaux), and were downloaded on 2024-10-19. This database feeds into the French National Inventory of Natural Heritage (SINP), supported by the French Ministry of the Environment. Opportunistic records are presence-only data from citizen science programs and surveys of specific areas (e.g., a natural reserve, a golf course) that do not follow a presence/absence or rigorous transect protocol. The inclusion of presence-only yet professional surveys of various locations implies that the dataset is therefore not necessarily biased towards high-richness or high-abundance areas. Standardized surveys were carried out along transects (as in the European Butterfly Monitoring Scheme), yet absences were not systematically recorded during the application of this method.

The dataset amounts a total of 298,389 valid records of 200 butterfly taxa along non-winter months (10, begin February-end November) for 24 years. The administrative region of Nouvelle-Aquitaine has 90,290 $1 \times 1$ km cells when represented as a grid. Butterfly records were allocated to the cells that comprised their original data types (e.g, points, transects). Our primary focus was on six species that are well-reported and vary in rarity as well as habitat specialization: \textit{Polyommatus icarus} (Rottemburg, 1775), \textit{Lycaena dispar} (Haworth, 1803), \textit{Maniola jurtina} (Linnaeus, 1758), \textit{Coenonympha oedippus} (Fabricius, 1787), \textit{Euphydryas aurinia} (Rottemburg, 1775), and \textit{Lycaena phlaeas} (Linnaeus, 1761). The data include 59,698 records of these six species (\textit{P. icarus}: 12,052, \textit{L. dispar}: 3,106, \textit{M. jurtina}: 17,465, \textit{C. oedippus}: 10,700, \textit{E. aurinia}: 6,982, \textit{L. phleas}: 9,378 records).

Data types are rather varied in this database and do not always fall easily into a ``standardized'' vs ``opportunistic'' dichotomy, so pooling all data sources into a single format and using occupancy models was a sensible option for modeling these data \citep{fletcherjr2019practical}, as opposed to the integrated modeling of a small set of well-delineated data sources, which can be done in other cases \citep{isaac2020data}. The data pooling approach has been used to model occupancy of butterfly species in the UK and Netherlands using data with heterogeneity and size similar or even larger than ours \citep{vanstrien2013opportunistic, fox2015state, dennis2017efficient, boyd2023operational, diana2023fast,  dennis2024efficient}.

Non-detections (zeroes) and sampling effort are inconsistently recorded in our data. Thus, only presences were used to produce occupancy data for individual species, with detection of any species in the butterfly community considered as evidence that sites were surveyed, allowing to produce detection/non-detection histories \citep{kery2010site, vanstrien2013opportunistic}. This decision implies assuming that any butterfly record is informative about the existence of survey effort. For this approach to work, a large proportion of sample visits (50\% or even more) must originate from community surveys \citep{shirey2023occupancy}. We surpassed that percentage with our data, as  91.5\% of the projects ($n=461$ out of the 504 projects) recorded two or more butterfly species or taxa, and of these 74\% ($n=373$) recorded six or more species.

\begin{figure}[htbp]
   \centering
    \includegraphics[width=1\linewidth]{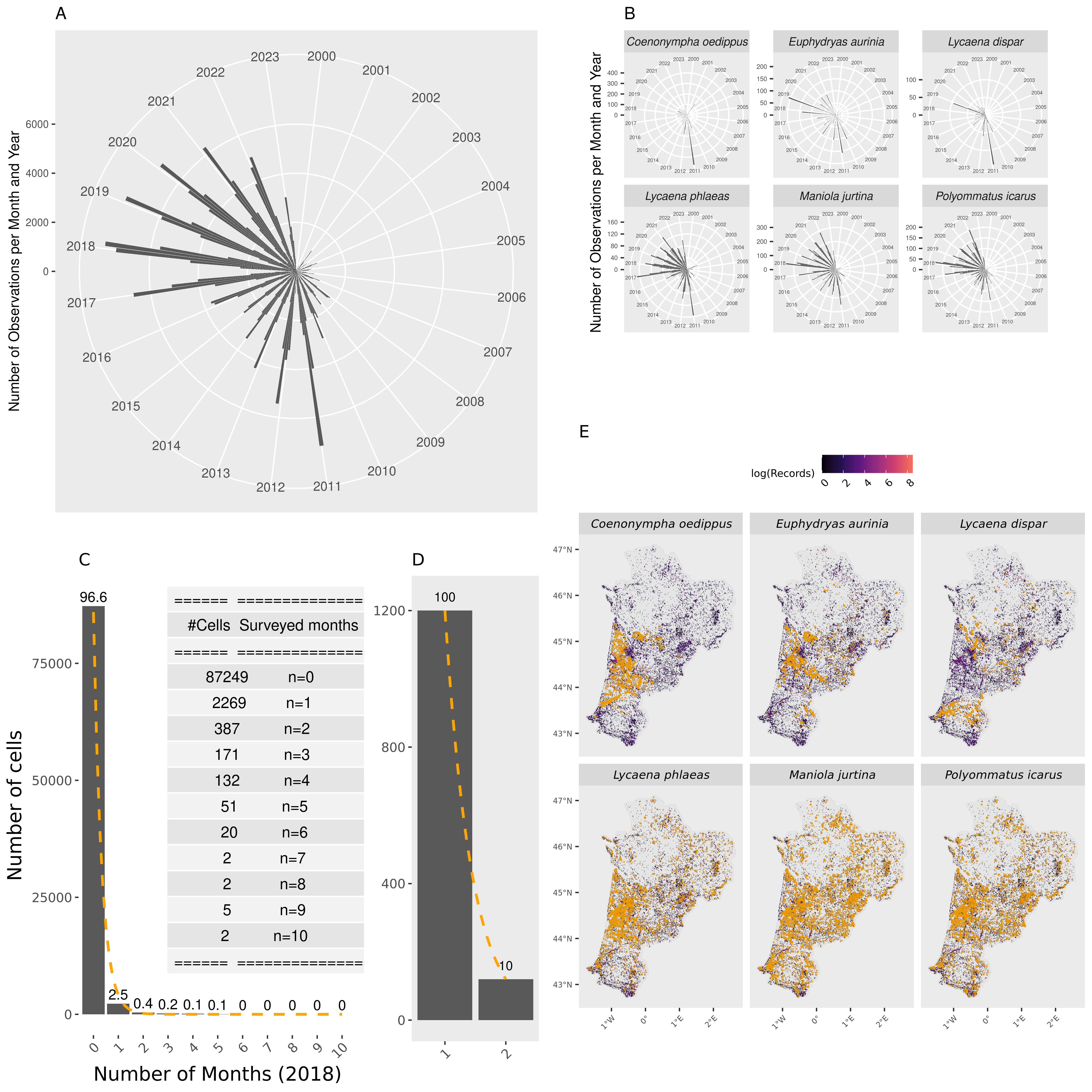}
    \caption{Description of the butterfly datasets and our focal species subsample. (A) Distribution of records and (B) species across months and years. (C) The distribution of records across cells and secondary occasions (months) of 2018. Percentages are shown over the bars, and the number of cells per number of surveyed months is shown in the inset table. (D) Data design of \citet{doser2024fractional} simulations, for comparison, where 100\% of the sites were surveyed at least once and 10\% were surveyed twice. (E) Focal species records (orange points) across the $T$=24 years of data, with effort (the total number of records over the years, at natural log scale) shown in the background. In (A) and (B), years are shown in the borders, and the vertical white bar marks June of each year (from 2000 to 2023). The orange curves in (C) and (D) depict the relationship between the number of cells and secondary occasions evaluated through a Poisson GLM with a quadratic function of secondary occasions fitted to the data. Butterfly data was taken from the  Nouvelle-Aquitaine Wildlife Observatory (FAUNA, \url{https://observatoire-fauna.fr/}). Grid cell resolution: $1 \times 1$ km.}
    \label{fig:butterfly-data}
\end{figure}

The resulting occupancy data -- an array of species encounter histories (detections and non-detections) aggregated at the level of sites ($I=90,290$ $1 \times 1$ km cells), primary occasions (year, $T=24$), and secondary occasions (survey months of each year, $J=10$) -- showed substantial heterogeneity. Sites had from 0 to 4,894 butterfly records in total (average $\pm$ SD: $3.3 \pm0.11$), with 19.6\% of the cells ($n=17,686$ cells) having at least one butterfly record and 80.4\% ($n=72,604$ cells) of the cells having zeroes for all 24 years. There was a very skewed distribution of records across sites, for all years (Fig. \ref{fig:butterfly-data}). Species records are well spread in space, especially for the common species \textit{Polyommatus icarus}, \textit{Lycaena phlaeas}, \textit{Maniola jurtina}, yet gaps and clusters of observations occur at specific locations and years. For instance, in 2018, the year with the largest number of records ($n=31,584$) in the dataset (Fig. \ref{fig:butterfly-data}A), 96.6\% of the cells were not sampled, 2.5\% were visited once, and 0.4\% were visited twice (Fig. \ref{fig:butterfly-data}C). This skewed distribution differs substantially from D\&S data design used to test the model (Fig. \ref{fig:butterfly-data}D). Most records were gathered at cells around Bordeaux during aural summer months (June, July) due to observers' preferences/constraints and butterfly phenology (Fig. \ref{fig:butterfly-data}A,B,E). Gaps in data and the skewed spatiotemporal distribution of surveys are hallmarks of opportunistic datasets that multi-season occupancy models must account for \citep{isaac2014statistics, kelling2019semistructured, isaac2020data, johnston2020estimating}. 

\subsection{Model}

Multi-season site occupancy models with spatial and temporal random effects consist of hierarchically related submodels that can be declined in the following manner, following \citet{doser2024fractional}. The first part of the model is an occupancy state process, where we model the latent occupancy state $z_{it}$ of a single species at $i=1, ..., I$ sites and $t=1, ..., T$ primary occasions. The state $z_{it}$ is drawn from a Bernoulli distribution depending on the probability of occupancy of the site $\psi_{it}$. $\psi_{it}$ is a function of the covariate(s) at the site and the primary occasion level $X_{it}$, and $\boldsymbol{\beta}$ is a vector of coefficients including the intercept and the slope representing the effect of the covariate $X_{it}$ on $\psi_{it}$ (eq. \ref{eq:occupancy_state}), with  $\mathbf{X}_{it}^T = [1 \; X_{it}]$ to keep with notations in \citet{doser2024fractional}. The model writes

\begin{equation}
\begin{split}
z_{it}|\psi_{it} \sim \mathcal{B}(\psi_{it}), \\
\text{logit}(\psi_{it}) = \mathbf{X}_{it}^T\boldsymbol{\beta} + \omega_{i} + \eta_{t}
\label{eq:occupancy_state}
\end{split}
\end{equation}

The spatial random effects $\omega_{i}$ are defined through a Gaussian process where for each vector of locations $\mathbf{s}$ we have
\begin{equation}
\boldsymbol{\omega}(\mathbf{s}) \sim \mathcal{N}(\mathbf{0},\boldsymbol{\Sigma}(\mathbf{D},\boldsymbol{\theta}))\label{eq:spatial-random-effects}
\end{equation}
where $\mathbf{D}$ is a distance matrix between all locations stored in $\mathbf{s}$. $\boldsymbol{\theta}$ includes the spatial decay $\phi$ and spatial variance $ \sigma^2$ that modulate the strength of spatial autocorrelation in continuous space in an exponential correlation model \citep{doser2024fractional}. The model applied in \cite{doser2024fractional} uses the Nearest Neighbor Gaussian Processes (NNGP) rather than the full Gaussian Processes to account for spatial autocorrelation \citep{datta2016hierarchical}, which makes it possible to estimate spatial autocorrelation parameters with large datasets \citep{doser2022spOccupancy}.

The temporal random effects $\eta_{t}$ follow a zero-mean AR(1) process with covariance
\begin{equation}
\text{Cov}(\eta_t,\eta_{t'}) = \sigma^2_T\times \rho^{|t-t'|},
\label{eq:temporal-random-effects}
\end{equation} where $\rho$ is the temporal autocorrelation and $\sigma^2_T$ the temporal variance. 

And the model is not complete without its observation process:
\begin{equation}
\begin{split}
y_{itj} | z_{it} \sim \mathcal{B}(z_{it} \times p_{itj}) ,\\ 
\text{logit}(p_{itj}) = \boldsymbol{v}_{itj}^T\boldsymbol{\alpha}
\label{eq:observation-process}
\end{split}
\end{equation}
where $\boldsymbol{\alpha}$ is a vector of coefficients, including the intercept and the slope that represent the effect of the covariate(s) $v_{itj}$ on $p_{itj}$ (eq. \ref{eq:observation-process}). The species encounter history $y_{itj}$ is then conditionally related to the latent occupancy $z_{it}$, meaning that for a truly occupied site and primary occasion $z_{it}=1$, the species will be detected in one individual secondary occasion with probability $p_{j}$ (eq. \ref{eq:observation-process}). If unoccupied $z_{it}=0$ then the species cannot be detected. This multi-season occupancy model is implemented in the \texttt{spOccupancy} package with function \texttt{stPGOcc} \citep{doser2022spOccupancy}.

\subsection{Simulated data design}

\begin{figure}[htbp]
    \centering
    \includegraphics[width=1\linewidth,trim={0 0 0 0},clip]{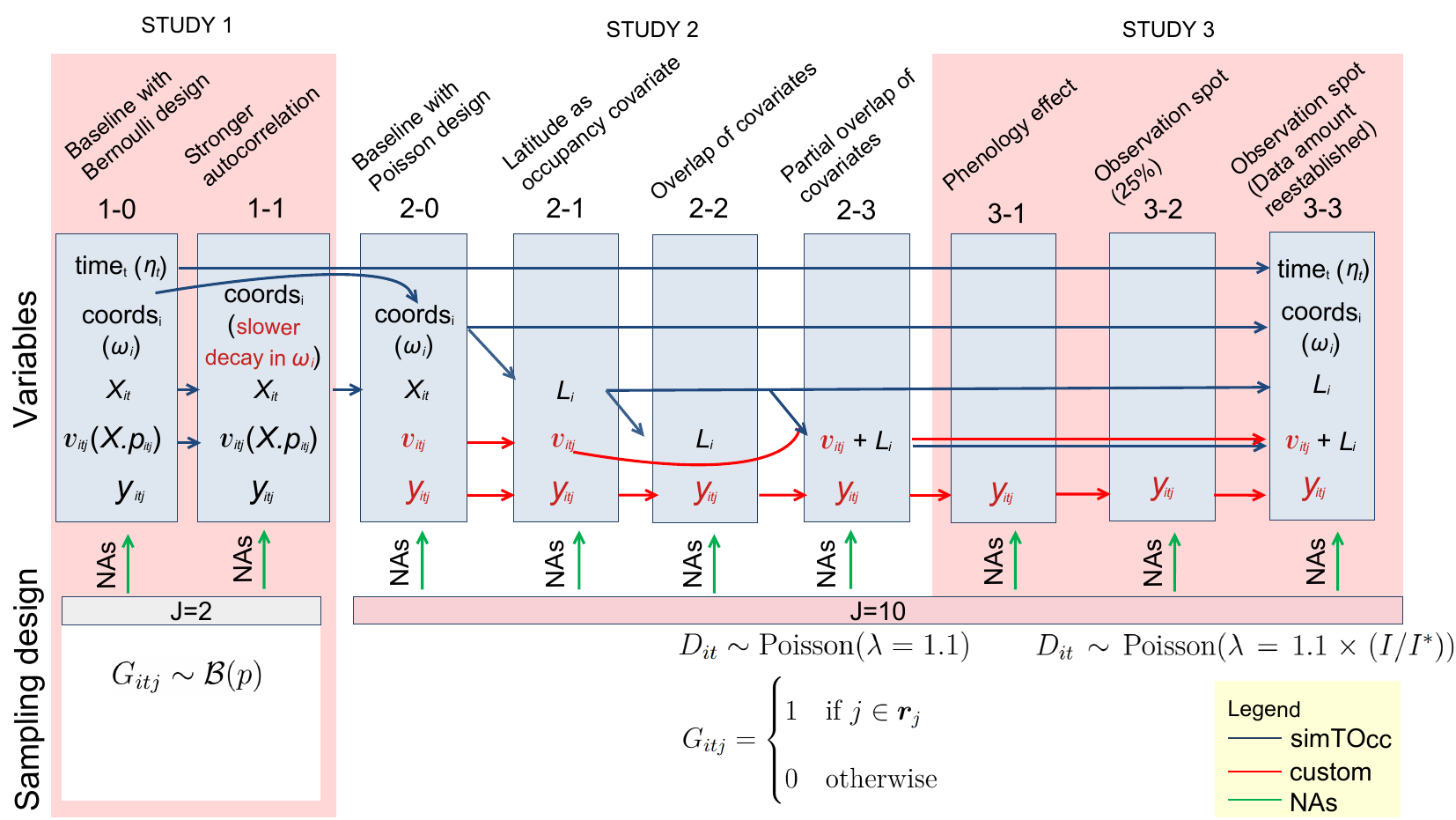}
    \caption{Simulation map. Study-scenario 1-0 represents the baseline design of \citet{doser2024fractional}. Temporal and spatial random effects are depicted by $\omega_i$ and $\eta_t$, respectively. Study-scenario 1-1 imposed a stronger spatial autocorrelation situation by setting slower spatial decays, with $\phi=1$ and $\phi=0.5$. Study 2-0 represents a shift from a Bernoulli to Poisson distribution to spread secondary occasions to sites. Covariates change from this scenario to the subsequent ones. In scenario 2-1 the random site-year covariate $X_{it}$ was replaced by site latitude $L_i$. Scenarios 2-2 and 2-3 represent different combinations of $L_i$ and $v_{itj}$. In 2-2 there is complete overlap of covariates between occupancy and detection models, and in 2-3 the overlap is partial. Study 3 incorporates temporal variation in detection (to represent phenology and observer sampling preferences for midseason, 3-1) and spatiotemporal structures in the occupancy data (phenology + observer preferences + observation spot, 3-2 and 3-3). Blue arrows show the data generated by the function \texttt{simTocc} (\texttt{spOccupancy} R package \citealt{doser2022spOccupancy}), and red arrows and text show data generated by customized functions. Blue arrows that overlap different boxes show the variables used in more than one study and scenario. Green arrows indicate the data in which missing observations $NAs$ were imputed according to the different sampling designs.}
    \label{fig:simulation-design}
\end{figure}

We ran three simulations studies to assess the parametric identifiability of the multi-season occupancy model with spatial and temporal random effects (Fig. \ref{fig:simulation-design}). 

The first simulation study was a replication of D\&S simulations (study 1, scenario 0). From a full dataset of $I=1,200$ sites, $T=10$ primary occasions and $J=5$ secondary occasions, D\&S created a heterogeneous design with up to $J=2$ secondary occasions within each primary occasion, using a two-group mixture of cells: 90\% of the sites were visited once within each primary occasion, 10\% were visited twice (Fig. \ref{fig:butterfly-data}D). Gaps within primary occasions were produced by a Bernoulli sampling design as follows

\begin{equation}
G_{itj} \sim \mathcal{B}(p)\label{eq:design-array-D&S}
\end{equation}
for all $j\in \{1,...,J\}$, where $G_{itj}$ is a sampling design array indicating which site will be sampled within each primary $t$ and secondary occasion $j$. In the design of D\&S, for $j=1$ the success probability was $p_1=1$ (all sites were sampled), for all $t \in \{1,..., T\}$. At $j=2$ the success probability was $p_2=0.1$ (Fig. A.1).

Recently, \citet{belmont2024spatiotemporal} showed that D\&S's model overestimates the spatial decay parameter $\phi$, which is likely caused by the use of sparse approximations for Gaussian processes (GP) in the nearest neighbor approach (based on \citealt{datta2016hierarchical}). To further evaluate if $\phi$ overestimation affects occupancy estimates, we built scenario 1 within study 1 (1-1, Fig. \ref{fig:simulation-design}) where data were simulated under stronger spatial autocorrelation/slower spatial decay levels ($\phi=0.5$ and $\phi=1$, Fig. A.2) than scenario 0.

The study 2 started with the replacement of a Bernoulli-distributed by a Poisson-distributed number of surveys to sites, in order to mimic the distribution of surveys in the butterfly data (Fig. \ref{fig:butterfly-data}). The Poisson design kept the total amount of data constant, which was achieved by summing the vector of probabilities of D\&S Bernoulli design $p_1+p_2 = 1+0.1 = 1.1$ and using it as the intensity parameter $\lambda$ of a Poisson distribution

\begin{equation}
D_{it} \sim \text{Poisson} (\lambda=1.1)
\label{eq:D_it}
\end{equation}
where $\boldsymbol{D}=(D_{it})$ is a matrix with the number of secondary occasions for $i \in \{1, ..., I\}, t \in \{1, ..., T\}$, with $D_{it} \in [0,J]$. The Poisson distribution yields a probability for a site to have zero surveys (spatial gap) in a given year of 33\% (Fig. A.3) and the probability of 0 surveys in $T=10$ years is $0.33^{10} = 1.6 \times 10^{-5}$, being low enough to be neglected (otherwise a truncated Poisson distribution might be used). 

$\boldsymbol{D}$ was then used to make a new sampling design array $\boldsymbol{G}$ with elements $G_{itj}$ (eq. \ref{eq:our-design-array}). To define which $j$ secondary occasions were sampled in each site and year, we spread $D_{it}$ across all $J$ in $G_{itj}$. The spreading was done with a random sampling algorithm without replacement and uniform sampling probabilities $\boldsymbol{p}=(p_1,..., p_J)=(0.1, ..., 0.1)$. The algorithm resulted in the vector $\boldsymbol{r}_j$ which matches the values in $D_{it}$. Then, $\boldsymbol{r}_j$ was used to indicate which $j$ will be sampled per site $i$ and primary occasion $t$, so that

\begin{equation}
\begin{split}
G_{itj} = 
\begin{cases} 
1 & \text{if } j \in \boldsymbol{r}_{j} \\
0 & \text{otherwise}\label{eq:our-design-array}
\end{cases}
\end{split}
\end{equation}

If $D_{i=1, t=1} = 5$, one possible result could be $G_{i=1,t=1,1<j<J } = (0,1,0,1,1,0,1,1,0,0,0)$. Thus, our study 2-scenario 0 (2-0) consisted in challenging the model with a more skewed distribution of the number of surveys, all else remaining equal to D\&S design (study 1-scenario 0, Fig. A.4, Fig. \ref{fig:simulation-design}). 

Using this Poisson design, we started to change the combinations of covariates in occupancy and detection models. In the previous scenarios, covariates affecting occupancy and detection in D\&S were fully random (uncorrelated) in space and time and with regard to each other. Then, in 2-1, we replaced $X_{it}$ by the scaled values of grid latitude $L_i$ in eq. \ref{eq:occupancy_state}, such that

\begin{equation}
\text{logit}(\psi_{it}) = \begin{bmatrix}1 & L_{i}\end{bmatrix}\begin{bmatrix}\beta_0 \\ \beta_1 \end{bmatrix} + \omega_{i} + \eta_{t} 
\label{eq:latitude-model-occ}
\end{equation}

The use of $L_i$ imposes a spatial structure in occupancy data (Fig. A.5). No change was made in the detection model. In (2-2), we replaced both $X_{it}$ and $v_{itj}$ by $L_i$ in eq. \ref{eq:occupancy_state} and \ref{eq:observation-process}, producing the overlap of covariates (Fig. A.6) already shown to challenge the performance of occupancy models \citep{lele2012dealing, peach2017single}. The model writes

\begin{equation}
\begin{split}
\text{logit}(\psi_{it}) =  \begin{bmatrix}1 & L_{i}\end{bmatrix}\begin{bmatrix}\beta_0 \\ \beta_1 \end{bmatrix} + \omega_{i} + \eta_{t} \\
\text{logit}(p_{itj}) =  \begin{bmatrix}1 & L_{i}\end{bmatrix}\begin{bmatrix}\alpha_0 \\ \alpha_1 \end{bmatrix}
\end{split}
\end{equation}
 
In (2-3), we added the observation-level covariate $v_{itj}$ to $L_i$ in the detection model, imposing a partial overlap of covariates between occupancy and detection models (Fig. A.7) with

\begin{equation}
\begin{split}
\text{logit}(\psi_{it}) = \begin{bmatrix}1 & L_{i}\end{bmatrix}\boldsymbol{\beta} + \omega_{i} + \eta_{t} \\
\text{logit}(p_{itj}) = \begin{bmatrix}1  & v_{itj} & L_{i} \end{bmatrix} \boldsymbol{\alpha}
\end{split}
\end{equation}

In our third study, we added temporal and spatial structures in occupancy data. In (3-1), we reformulated the sampling design (eq. \ref{eq:our-design-array}) to represent phenology and observer sampling preferences for midseason. We maintained $\lambda=1.1$ and $J=10$ secondary occasions, but used non-uniform probabilities in $\boldsymbol{p}_J$. This vector was obtained from a Gaussian function multiplied by a small noise $A$, centered at the peak of the surveyed occasion $\mu$ of each year

\begin{equation}
\boldsymbol{p}_J=\left(e^{A \times (-(\frac{j-\mu}{\sigma})^2)}\right)_{j \in \{1, ..., J\} }
\label{eq:gaussian-function}
\end{equation}
where $\mu=J/2$. $\sigma$ is the spread of the peak (set as $\sigma=J/4$), causing probability drops before and after the peak. The parameter $A$ is drawn as $A \sim \mathcal {N}(0,0.33^2)$, creating some variation around $\mu$. We then ranked $\boldsymbol{p}_J$ and selected its $k$ largest values, where $k$ is the number of secondary occasions in $D_{it}$. Then, we created a new sampling array $G_{itj}$ (eq. \ref{eq:our-design-array}). The resulting occupancy data (Fig. A.8) could represent for instance a univoltine butterfly displaying a single activity peak in the middle of the year \citep{bishop2013utility}.

In our scenario 3-2, we reformulated the Poisson sampling design to represent both temporal and spatial clustering of observations in the simulated data. We used eq. \ref{eq:gaussian-function} to obtain a site-wise sampling probability vector $\boldsymbol{p}_I=(p_1,p_2, ..., p_I)$ by setting $\mu=I/2$ (mid-latitude peak) and $\sigma=I/2$ (small spread). To create the observation spot in scenario 3-2, we ranked $\boldsymbol{p}_I$ and selected the 25\% ($I^*=300$) out of the $I$ sites with the largest probability values. Then, we created a new sampling design array $G_{itj}$, which depicted a mid-latitude clustering of sampled sites in each year (Figs. A.9).

Creating this spatial spot with a Poisson intensity $\lambda=1.1$ caused a decrease in the amount of data from 3-1 to 3-2. To reestablish the amount of data used in all previous scenarios, we created scenario 3-3 where we increased the intensity parameter used in eq. \ref{eq:D_it} to $D_{it} \sim \text{Poisson} (\lambda=1.1 \times (I/ I^*))$. This means that we distributed the same amount of data used in scenarios 1-0 through 3-1 within the spot (Fig. A.10). Then, the sampling design array $G_{itj}$ was recreated to depict the mid-latitude observation spot (Figs. A.10). Since spatial and temporal gaps were produced in the last two scenarios, out-of-sample predictions from the model were necessary to estimate occupancy probability for unsampled sites and years.

\subsection{True parameter values, MCMC settings, and Software}

Pairwise combinations of the values of $\phi$, $\sigma^2$, $\rho$ and $\sigma^2_T$ (Table \ref{table:parameter-description}) resulted in 16 analyzed sub scenarios of spatial and temporal autocorrelation within each study and sampling-design (Table A.1). We simulated 100 datasets under these 16 sub scenarios within each study and data design (Fig. \ref{fig:simulation-design}), yielding the analysis of 14,400 datasets.

For analyzing each dataset, we used 25,000 iterations in each one of three parallel MCMC chains, a burn-in phase of 15,000 iterations, and thinning each 10 iterations, yielding 3,000 posterior distribution samples per parameter. These samples were subsequently used to obtain point estimates (averages) used in statistical analyses. We used five neighbors in the nearest neighbor Gaussian Process (NNGP) approximation. 

We used the following weakly informative priors to fit the models: $\phi \sim U(3,60)$, $\boldsymbol{\alpha},~ \boldsymbol{\beta}\sim\mathcal{N}(\mu=0,\sigma^2=2.72)$, $\sigma^2\sim 1/\Gamma(a=2,b=1.5)$, $\sigma{^2}_{T} \sim 1/\Gamma(a=2,b=1)$, $\rho \sim U(-1,1)$. The same priors were used in simulations and empirical data analyses.

All simulations were done using functions available in the R package \texttt{spOccupancy} version 0.7.6 \citep{doser2022spOccupancy} and using our own custom codes. We used R version 4.4.1 \citep{rcoreteam2024R}. Figures and maps were produced using the R package \texttt{ggplot2} \citep{wickham2016ggplot2} and \texttt{sf} \citep{pebesma2023spatial}. All code and information about package versions are available on our GitHub page (see Data Availability Statement).

\begin{table}[htbp]
\centering
\caption{True parameter values used in the simulations. These are the same as those used by \cite{doser2024fractional}. Asterisks highlight the values of the spatial decay $\phi$ that were changed to $\phi=0.5$ and $\phi=1$ in Study 1-Scenario 1. Parameter values separated by `/' represent low and high values used in spatial and temporal autocorrelation scenarios.}
\begin{tabular}{|c|c|c|} 
 \hline
Parameter & True value & Description \\ 
\hline
\(\beta_0\) & 0 & Intercept of the occupancy model (logistic scale)\\
\(\beta_1\) &  0.5 & Effect of $X_{it}$ or $L_{i}$ on $\psi_{it}$\\
\(\sigma^2\) & 0.3/1.5 & Spatial variance used in $\omega_i$\\
\(\phi\) & 3.75*/15* & Spatial decay used in $\omega_i$\\
\(\rho\) & 0.5/0.9 & Temporal correlation used in $\eta_t$\\
\(\sigma^2_T\) & 0.3/1.5 & Temporal variance used in $\eta_t$\\
\(\alpha_0\) & 0 & Intercept of the detection model (logistic scale)\\
\(\alpha_1\) & -0.5 & Effect of $v_{itj}$ on detection\\
\(\alpha_2\) &  -0.5 & Effect of $L_{i}$ on detection\\ 
\hline
\end{tabular}
\label{table:parameter-description}
\end{table}

\subsection{Identifiability assessment}

As a reminder, let us state that model identifiability refers to the ability to uniquely determine the values of model parameters from the available data under parametric model assumptions \citep{gimenez2004methods}. A model is globally identifiable if there is a one-to-one correspondence between its parameters and the model \citep{gimenez2004methods, cole20202parameter}. In a locally identifiable model, only a few parameter values can produce the observed data with the same likelihood. 

As the framework used is Bayesian, and we still wish to evaluate estimator properties in a frequentist sense, we use the posterior distribution mean across MCMC draws \cite[p. 127-128]{cole20202parameter}. The distribution of posterior means across simulated datasets is therefore used to diagnose parameter identifiability. 

We initially used scatter plots to assess bias on point estimates of $\hat{\psi_{it}}$ relative to the true $\psi_{it}$ (as done in D\&S), for each study, scenario, and sub-scenario of spatial and temporal autocorrelation. Bias on occupancy probability was diagnosed whenever the obtained relationship deviated from a 1:1 relationship, which represents the perfect match between $\hat{\psi_{it}}$ and $\psi_{it}$. In addition to the scatter plots, we made spatial maps of $\hat{\psi_{it}}$, $\psi_{it}$ and of their differences, enabling the identification of bias in space. We did these maps for a single simulated dataset under the two most extreme sub scenarios of spatial and temporal autocorrelation: low parameter values ($\phi=3.75$ [or $0.5$ in study-scenario 1-1], $\sigma^2=0.3$, $\rho=0.5$, $\sigma^2_T=0.3$) and high parameter values ($\phi=15$ [or $1$ in study-scenario 1-1], $\sigma^2=1.5$, $\rho=0.9$, $\sigma^2_T=1.5$), see Table \ref{table:parameter-description} for a description of all true parameter values).

Scatter plots were used to evaluate the relationship between the true and estimated product of occupancy and detection probability. There may be scenarios where $\psi$ and $p$ cannot be estimated separately but their product (apparent occupancy probability) can be estimated, therefore diagnosing a problem of identifiability \citep{gimenez2004methods, cole20202parameter}. Thus, we obtained the true $\psi_{it} \times p_{it}$ and the estimated $\hat{\psi_{it}} \times \hat{p_{it}}$ for each simulated dataset. Note that the detection probability was aggregated at site and time level by calculating its average across the $J$ survey occasions.

Barplots were used to evaluate how the mean squared errors ($\text{MSE}(\hat{\psi_{it}})=\mathbb{E}[(\hat{\psi_{it}} - \psi_{it})^2]$) of the occupancy estimator varied across studies and scenarios. The MSE was also calculated for the estimators of $\phi$ and $\rho$.

Contour plots were used to evaluate linkages between pairs of parameters found together in the models. These combinations were i) model intercepts: $\hat{\beta_{0}}$ vs $\hat{\alpha_{0}}$; ii) intercepts and slopes ($\hat{\beta_{0}}$ vs $\hat{\beta_{1}}$, $\hat{\alpha_{0}}$ vs $\hat{\alpha_{1}}$ and $\hat{\alpha_{0}}$ vs $\hat{\alpha_{2}}$, iii) detection slopes $\hat{\alpha_{1}}$ vs $\hat{\alpha_{2}}$, iv) spatial autocorrelation coefficients $\hat\phi$ vs $\hat\sigma^2$, and v) temporal autocorrelation parameters $\hat\rho$ vs $\hat\sigma^2_T$. Results were shown for the scenarios of high temporal correlation and variance ($\rho=0.9$ and $\sigma^2_T=1.5$), which is the case where there is high sharing of temporal information and, as such, temporal random effects could contribute more to model identifiability and inference. Results for lower temporal autocorrelation levels are shown in the Online Supporting Information.

In each contour plot, a single region of high density of point estimates is expected for a globally identifiable model, with the true value of each parameter centered inside the high-density region \citep{cole20202parameter}. More than one high-density region can indicate local identifiability, and an elongate-shaped density or no density at all can represent an identifiability issue \citep{cole20202parameter}. Densities were estimated using a two-dimensional Gaussian kernel density estimator of the \texttt{MASS} R package \citep{venables2002modern}, and projected across parameter combinations using \texttt{ggplot2} \citep{wickham2016ggplot2}. 

\subsection{Misspecification assessment}

Model misspecification represents the ways that models might fail to represent the true data generating process. Among the causes of misspecification are the omission of important covariates, wrong forms of covariates in the model, the use of wrong link functions and unmet parametric assumptions, among others \citep{stoudt2023nonparametric, doser2024fractional}. 

D\&S evaluated model robustness to misspecification by considering cases where the link function used to simulate the true $\psi_{it}$ and $p_{itj}$ differed from the link function used to estimate the parameters. For example, the data was simulated with probit or linear link function, but the fitted model had a logit-link function. We hereby evaluated misspecification by using the probit as the data generating function, and the logit as the model link function. This is a relatively mild case of misspecification, which involves a data-generating model that is not too far from the fitted model. We assessed model misspecification in four of our scenarios: 1-0, 2-2, 3-2 and 3-3 (Fig. \ref{fig:simulation-design}). Twenty datasets were simulated per study and scenario, yielding the additional analysis of 320 datasets (20 datasets $\times$ 16 autocorrelation levels). The same scatter plots described above were used to evaluate whether model misspecification affects the relationship between $\psi_{it}$ and $\hat{\psi_{it}}$.

A related concept is non-parametric model identifiability \citep{stoudt2023nonparametric}, but it remains slightly different from a robustness check to misspecification because it involves comparing whether identifiability is achieved in a parametric model vs in a larger nonparametric model which contains the parametric one as a special case. 

\subsection{Empirical data analysis}

We fitted the occupancy model to the encounter history of the common blue \textit{Polyommatus icarus} and the large copper \textit{Lycaena dispar}. We used the full Nouvelle-Aquitaine dataset (results presented in the Supporting Information) as well as a subset of the data from a buffer zone of 10 km$^2$ around the city of Bordeaux where sampling effort is larger (hereafter referred to as ``greater Bordeaux area'', and presented in the main text). This subset comprised 1,346 $1 \times 1$ km$^2$ cells, of which 702 had at least one butterfly record over the 24 years of data. Data from these 702 sites were used to fit the model, and predictions from the model were obtained for the remaining 644 cells without butterfly records. The greater Bordeaux area includes a similar number of sites to simulations, and comprises environmental and sampling effort gradients that might influence species occupancy and detection. We used a set of general covariates in our models (latitude, altitude, land cover, among others) which likely left some species occupancy variation unexplained, a situation in which spatial and temporal autocorrelation could improve model performance. Furthermore, it makes sense to expect autocorrelation in our data due to butterfly metapopulation dynamics (colonizations-extinctions over time) between neighboring sites \citep{hanski1996quantitative}. A full description of the covariates used in the models can be found in the Supporting Information C.

We anticipated an urban-countryside trend with lower occupancy in more heavily urbanized areas (center of the buffer) for the common blue. We also expected an east-west trend in the predicted distribution of the common blue, with lower occupancy probability in the west where less favorable (forested) habitats predominate. For the large copper, we expected higher occupancy around humid areas and rivers, more numerous in the northern part of the greater Bordeaux area. After accounting for imperfect detection, we expected stable occupancy trends over time for both species.

Models were built with 15 neighbors in the nearest neighbor Gaussian Process (NNGP) approximation, thus capturing fine-scale spatial autocorrelation. Prior-posterior overlap was used to evaluate model extrinsic identifiability front to real data \citep{cole20202parameter}. If substantial prior-posterior overlap exists, then the prior drives the posterior distribution and the data may have little influence on the results, whereas a small overlap means the data were informative enough to overcome prior's influence. The MCMC settings were 100,000 iterations each one of three MCMC chains, burn-in of 98,000 iterations, batch length of 100 iterations, and thinning each 5 iterations. These settings yielded 1,200 posterior distribution draws per parameter, and were used to make predictions and inference on butterfly occupancy and detection. The percentage of prior-posterior overlap was calculated using the R package \texttt{MCMCvis} \citep{youngflesh2018mcmcvis}. 

The mapped site-level occupancy probability for each species and posterior distribution draw was the averaged occupancy across years $\mathbb{E}(\hat{\psi_{i}}) = {\frac{1}{T}\sum_{t=1}^{T}~\hat{\psi_{it}}}$; subsequently we take the average across draws. The mapped spatial random effect $\mathbb{E}(\hat{\omega_{i}})$ was the average of $\hat{\omega_{i}}$ across the posterior distribution draws. The yearly occupancy trends for each species and posterior distribution draw was $\mathbb{E}(\hat{\psi_{t}}) = {\frac{1}{I}\sum_{i=1}^{I}~\hat{\psi_{it}}}$, the summed occupancy across sites relative to the total number of sites $I$; subsequently we take the average across draws. Estimated yearly occupancy was compared with the naive yearly occupancy, defined as the number of cells with detection relative to the total number of sampled cells per year. Finally, variation of detection probability across survey months was obtained by making predictions from the detection model using the estimated regression parameters for each posterior distribution draw. The point estimate (average trend) and 95\% Credible Intervals were calculated using all 1,200 posterior distribution draws.

\subsection{Sensitivity analyses applied to simulated data}

We evaluated the performance of another model with spatially uncorrelated random effects and random walk prior for temporal autocorrelation \citep{outhwaite2018prior}. This model is useful to estimate species occupancy trends based on large and sparse data \citep{outhwaite2019annual, boyd2023operational}, and it is simpler than the occupancy models shown above because it has only temporal autocorrelation. The model was fitted to 640 simulated occupancy datasets (16 sub scenarios $\times$ 42 simulation runs) created by imposing the conditions of study 2-scenario 1 (occupancy and detection models had different covariates -- $L_i$ and $v_{itj}$, respectively). A full description of the model and associated results is shown in Supporting Information F stored on our GitHub page.

Still with simulated data, we fitted an occupancy model with NNGP=15 spatial neighbors (rather than five) to data simulated under study-scenario 3-2 and 3-3. Thus, model specifications became more similar between simulations and empirical analyses. Finally, we evaluated model performance by fitting the model to occupancy data from a larger observation spot. This spot was created by selecting 50\% ($I^*=600$) out of the $I$ sites with the largest probability values, producing less sparse data as compared to scenario 3-2.

\subsection{Sensitivity analyses applied to empirical data}

The \texttt{stPGocc} model was fitted to the data of the four remaining species: the false ringlet \textit{Coenonympha oedippus} (Fabricius, 1787), the marsh fritillary \textit{Euphydryas aurinia} (Rottemburg, 1775), the small copper \textit{Lycaena phlaeas} (Linnaeus, 1761), and the meadow brown \textit{Maniola jurtina} (Linnaeus, 1758). These analyses were done at the greater Bordeaux scale, and we used weakly informative priors for $\phi$ ($\phi \sim U(3,60)$). These results are presented in Supporting Information G (see the Data Availability Statement).

In another analysis, we tested the sensitivity of the \texttt{stPGocc} model results to the prior chosen for $\phi$. Here, we replaced the weakly informative prior $\phi \sim U(3,60)$ by a more informative one $\phi \sim U(0.5,3)$, as in \citet{bajcz2024within}. A substantial prior-posterior overlap is expected for $\hat\phi$ because the informative prior constrains the MCMC sampler on specific regions of the parameter's distribution \citep{cole20202parameter}.

Additionally, we tested the sensitivity of the results (in particular the autocorrelation parameters) to models fitted to butterfly data covering the full Nouvelle-Aquitaine region. Occupancy data from 15 years and $I$=17,250 $1 \times 1$ km cells were used to fit the model. Here, we added the linear effect of natural grassland cover (inexistent within the greater Bordeaux area) as common blue occupancy predictor. Analyses were done with NNGP=15 spatial neighbors, and weak and informative priors for $\phi$. Out-of-sample predictions from the model were made for the full Nouvelle-Aquitaine dataset. To avoid RAM constraints, we used 50,000 iterations each one of three MCMC chains, burn-in of 48,000 iterations, batch length of 100 iterations, and thinning each 5 iterations. These settings yielded 1,200 posterior distribution draws per parameter. To avoid RAM errors, we made predictions using small groups of cells (61 groups of 1,500 cells) one at a time, and obtained the averages $\mathbb{E}(\hat{\psi_{i}})$ and $\mathbb{E}(\hat{\omega_{i}})$ per site and across posterior distribution draws. Subsequently, these values were projected onto maps.

\section{Results}

\begin{figure}[htbp]
    \centering
    \includegraphics[width=0.8\linewidth]{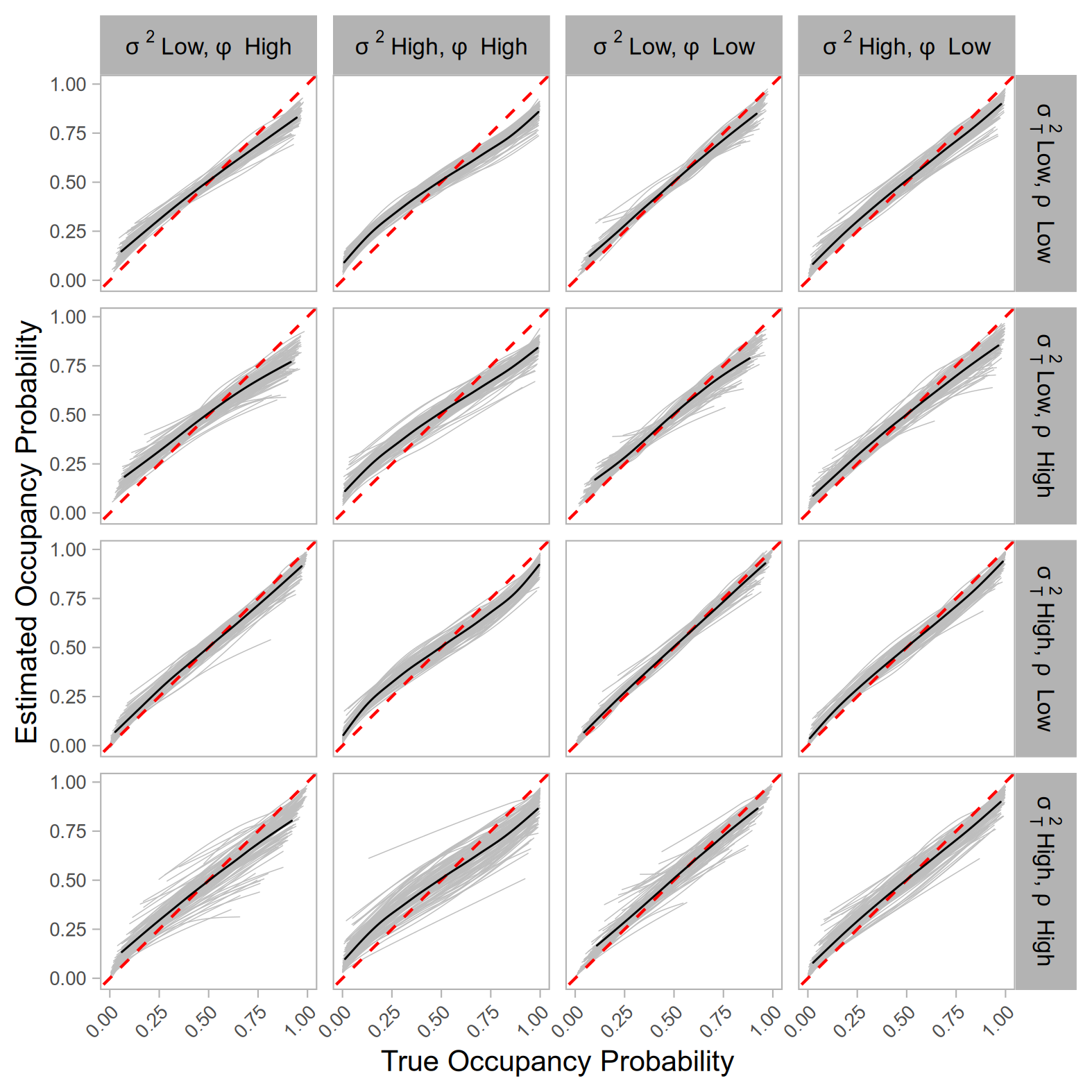}
    \caption{Relationship between the true $\psi_{it}$ (x-axis) and the estimated $\hat{\psi_{it}}$ (y-axis) occupancy probability when there was an overlap of covariates in occupancy and detection models (study 2-scenario 2). Each gray line represents the locally estimated scatterplot smoothing (LOESS) relationship between $\psi_{it}$ and $\hat{\psi_{it}}$ per simulated dataset. The black line depicts the averaged relationship across the 100 simulated datasets, and the red dashed line depicts a 1:1 relationship. Spatial and temporal autocorrelation sub scenarios are represented along columns and rows, respectively.}
    \label{fig:study1-scenario2-results}

\end{figure}

The tight relationship between true site occupancy $\psi_{it}$ and estimated site occupancy $\hat{\psi_{it}}$ found by \citet{doser2024fractional}, whose original relationship in shown in Fig. B.1, changed little across our scenarios of high spatial autocorrelation (study 1-scenario 1, Fig. B.2), skewed distribution of surveys (2-0, Fig. B.3), latitude as occupancy predictor (2-1, Fig. B.4), total overlap (2-2, Fig. \ref{fig:study1-scenario2-results}) and partial overlap of covariates (2-3, Fig. B.5), and phenology + observer sampling preferences (3-1) (Fig. B.6). In these situations, the larger deviations from the truth occurred for the scenarios with high spatial decay $\phi$, high spatial variance $\sigma^2$, high temporal correlation $\rho$, and high variance $\sigma^2_T$ (results not different from D\&S). For all scenarios, there was a subtle trend for overestimating occupancy when it was truly low (the estimated line was above the 1:1 relationship), and underestimating occupancy when it was truly high (the estimated line was below the 1:1 relationship). 

\begin{figure}[htbp]
    \centering
    \includegraphics[width=0.8\linewidth]{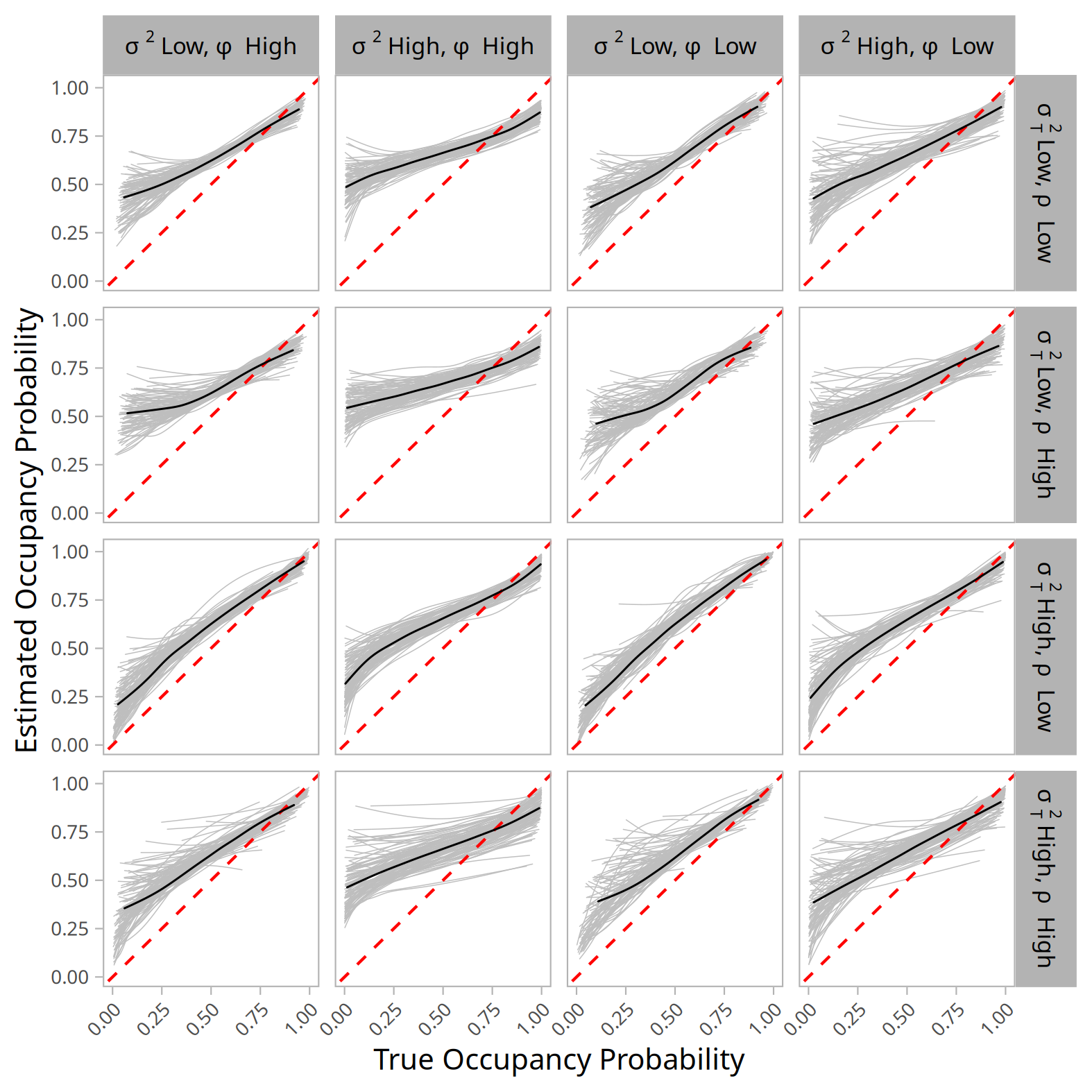}
    \caption{Relationship between the true $\psi_{it}$ (x-axis) and the estimated $\hat{\psi_{it}}$ (y-axis) occupancy probability when there was an partial overlap of covariates in occupancy and detection models, and temporal and spatial clustering of observations (study 3-scenario 3). Each gray line represents the locally estimated scatterplot smoothing (LOESS) relationship between $\psi_{it}$ and $\hat{\psi_{it}}$ per simulated dataset. The black line depicts the averaged relationship across the 100 simulated datasets, and the red dashed line depicts a 1:1 relationship. Spatial and temporal autocorrelation sub scenarios are represented along columns and rows, respectively.}
    \label{fig:study3-scenario3-results}

\end{figure}

Considerable deviations from the perfect 1:1 relationship between $\psi_{it}$ and $\hat{\psi_{it}}$, and substantially larger mean squared errors (MSE), occurred in the study-scenario 3-2 (Fig. B.7) and 3-3 (Fig. \ref{fig:study3-scenario3-results}). In these cases, $\hat{\psi_{it}}$ was biased high across most of the true $\psi_{it}$ range (Fig. \ref{fig:study3-scenario3-results}, Fig. B.8). The bias occurred for all spatial and temporal autocorrelation levels, being less severe when spatial variance was low and temporal variance was high. The pattern for $\hat{\psi_{it}}$ (Fig. \ref{fig:study3-scenario3-results}) resembled the pattern in the spatial random effect $\hat{\omega_i}$, which was biased in itself (Fig. B.9). There was an underestimation of $\hat{\phi}$ when it should be high $\phi=15$, indicating that estimated spatial autocorrelation was higher than it should be (Fig. B.10). When the true spatial decay was low $\phi=3.75$, the spatial decay estimates $\hat{\phi}$ were higher than they should be and uncertain for scenario 3-2 and also 1-1 (Fig. B.10). 

Similar bias on $\hat\psi_{it}$ was evidenced in our assessment of model misspecification (Supporting Information E, Figs. E.1-E.4) and evaluation of $\psi_{it}\times p_{it}$ (Figs. E.5-E.13). In addition, we found that the product of $\psi_{it}\times p_{it}$ had some bias from studies-scenarios 2-2 to 3-3 (Figs. E.9-E.13), most notably when the spatial variance level was low.

When mapped in space, the regions of truly high or low occupancy were visible across studies and scenarios (except for 3-2 and 3-3) (Figs. B.11-12). We noted that the estimates of occupancy $\hat{\psi_{it}}$ were less nuanced (more homogeneous maps with less fine-grained patterns) than the true occupancy after we started to change the combination of covariates. However, in the last two scenarios, occupancy overestimation was widespread in space (Figs. B.11-12). Despite this challenging condition, the model get closer to the true occupancy within the observation spot (mid-latitude) when $\phi$ was low (high spatial autocorrelation, Fig. B.11). Overall, out-of-sample predictions of occupancy did not match the true occupancy, especially for low occupancy areas (bottom of the simulated landscape, Figs. B.11-12).

Contour plots of combinations of model intercepts ($\hat{\beta_{0}}$ and $\hat{\alpha_{0}}$, at logistic scale) showed elongated densities across the $\hat{\beta_{0}}$ axis (Figs. B.13-B.14), especially when temporal correlation and variance were both high (Fig. B.14). Despite the large spread of $\hat{\beta_{0}}$ estimates, the true parameter values were positioned inside the region of high density of point-estimates. The exceptions were scenarios 3-2 and 3-3. In these cases, there were precise (low spread of point estimates) yet biased high $\beta_{0}$ estimators: the true value was outside the high density region (Figs. B.13-14). To sum up, overall, there are imprecise but unbiased estimates of the yearly average site occupancy before imposing scenarios 3-2 and 3-3. When these scenarios were considered, site-level occupancy estimates became biased in the sense that they were closer to yearly average site occupancy ($\hat{\beta_0}$) than they should have been. Average site occupancy was itself overestimated, so that local occupancy at grid cell level was overestimated as well (Figs. B.11-12).

For the occupancy model intercept and slope ($\hat{\beta_{0}}$ and $\hat{\beta_{1}}$), contour plots showed that the replacement of a spatiotemporal covariate $X_{it}$ by a site-level covariate $L_i$ increased variation around $\beta_{1}$, especially when spatial variance was high (Figs. B.15-B.16). Variation around $\hat{\beta_{0}}$ and $\hat{\beta_{1}}$ was larger when temporal correlation and variance were truly high (Fig. B.16). Nonetheless, the estimator was more severely biased in the scenarios 3-2 and 3-3 (Figs. B.15-16). No issue was found for the combinations of $\hat{\alpha_{0}}$ and $\hat{\alpha_{1}}$ (Figs. B.17-18). Scenario 3-2 showed a subtle bias and imprecision for estimates of $\hat{\alpha_{2}}$ (effect of $L_{i}$ on detection) (Figs. B.19-20).

\begin{figure}[htbp]
    \centering
    \includegraphics[width=1\linewidth]{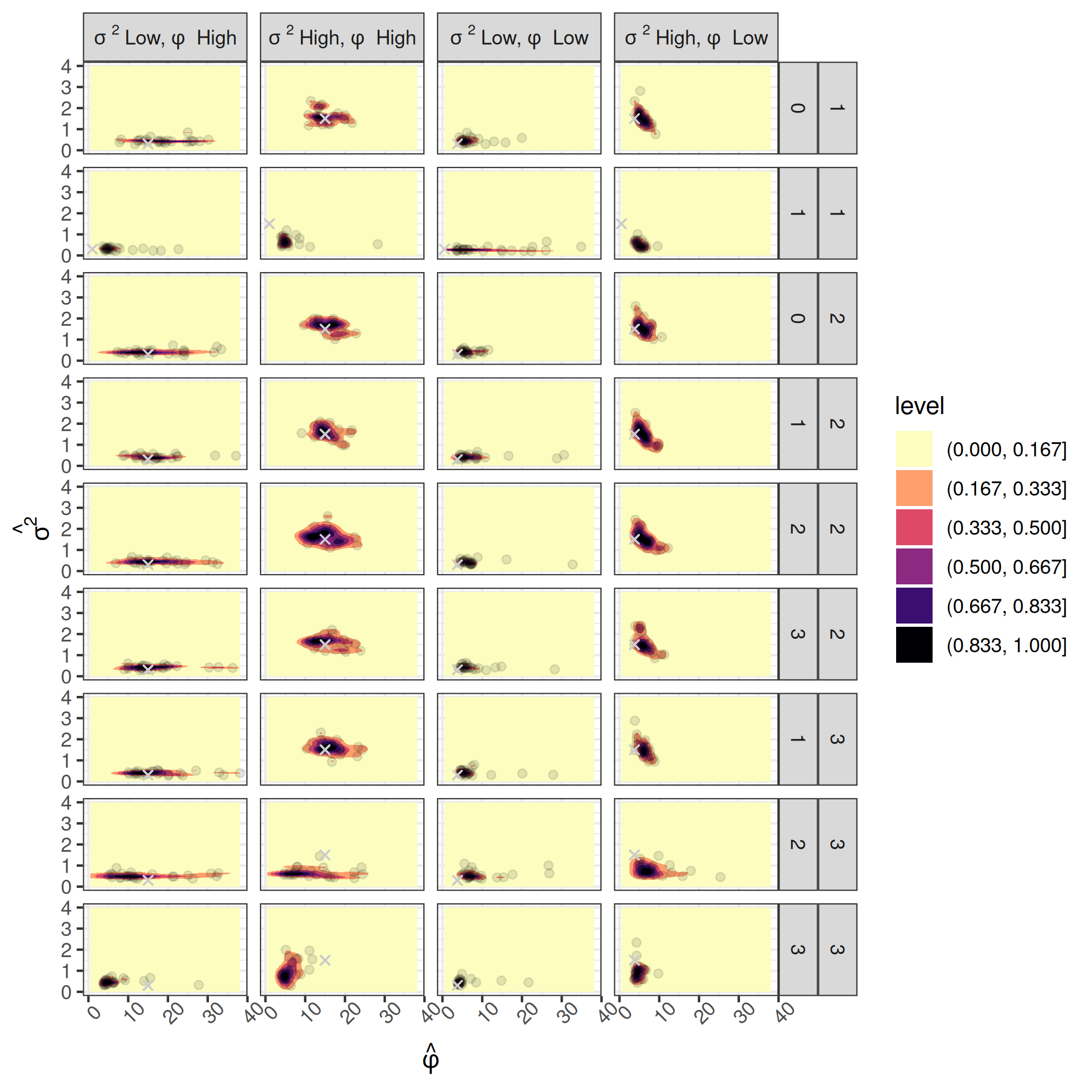}
    \caption{Contour plots showing the combination of point estimates of the spatial decay ($\hat{\phi}$, x-axis) and variance ($\hat{\sigma^2}$, y-axis). Studies and data design scenarios are shown across the facet rows, and spatial autocorrelation sub scenarios shown along facet columns. These results were produced by the sub scenarios with high temporal correlation $\rho=0.9$ and temporal variance $\sigma^2_T=1.5$. The density is given by 100 simulation runs per dataset. `level' in the legend depicts the discretized proportion of runs within each contour plot region. The gray crosses depict the true parameter values. Low $\phi=3.75$ and high $\phi=15$, except for study-scenario 1-1 where low $\phi=0.5$ and high $\phi=1$. Low $\sigma^2=0.3$ and high $\sigma^2=1.5$.}
    \label{fig:contour-phi-sigmasq}
\end{figure}

The contour plots with combinations of point estimates of spatial autocorrelation and variance parameters $\hat{\phi}$ and $\hat{\sigma^2}$ showed an elongated shape when the spatial decay $\phi$ was high and the variance $\sigma^2$ was low (Fig. \ref{fig:contour-phi-sigmasq}). When spatial decay and variance were both truly high, the estimated values of $\hat\phi$ and $\hat\sigma^2$ were generally inside of one of the high density regions, with exception of study-scenario 1-1, 3-2 and 3-3, when biased estimators were recovered (Fig. \ref{fig:contour-phi-sigmasq}, Figs. B.21-B.22). For the other two autocorrelation levels -- low $\sigma^2$ - low $\phi$, and high $\sigma^2$ - low $\phi$, shown in the two last columns of Fig. \ref{fig:contour-phi-sigmasq} and Fig. B.21 -- there was a subtle bias in the estimation of $\phi$ (all scenarios except 3-2) and underestimation of $\sigma^2$ (scenarios 1-1, 3-2 and 3-3). The MSE of $\hat{\phi}$ was high overall, especially when the spatial decay was truly high and scenarios 1-1, 3-2 and 3-3 were applied to data (Fig. B.22). Nonetheless, the lowest levels of error on $\phi$ estimator were observed when true spatial decay and variance were both high ($\phi=15$, $\sigma^2=1.5$) and under scenarios 1-0 and 2-0 to 3-1 (Fig. B.22).

\begin{figure}[htbp]
    \centering
    \includegraphics[width=1\linewidth]{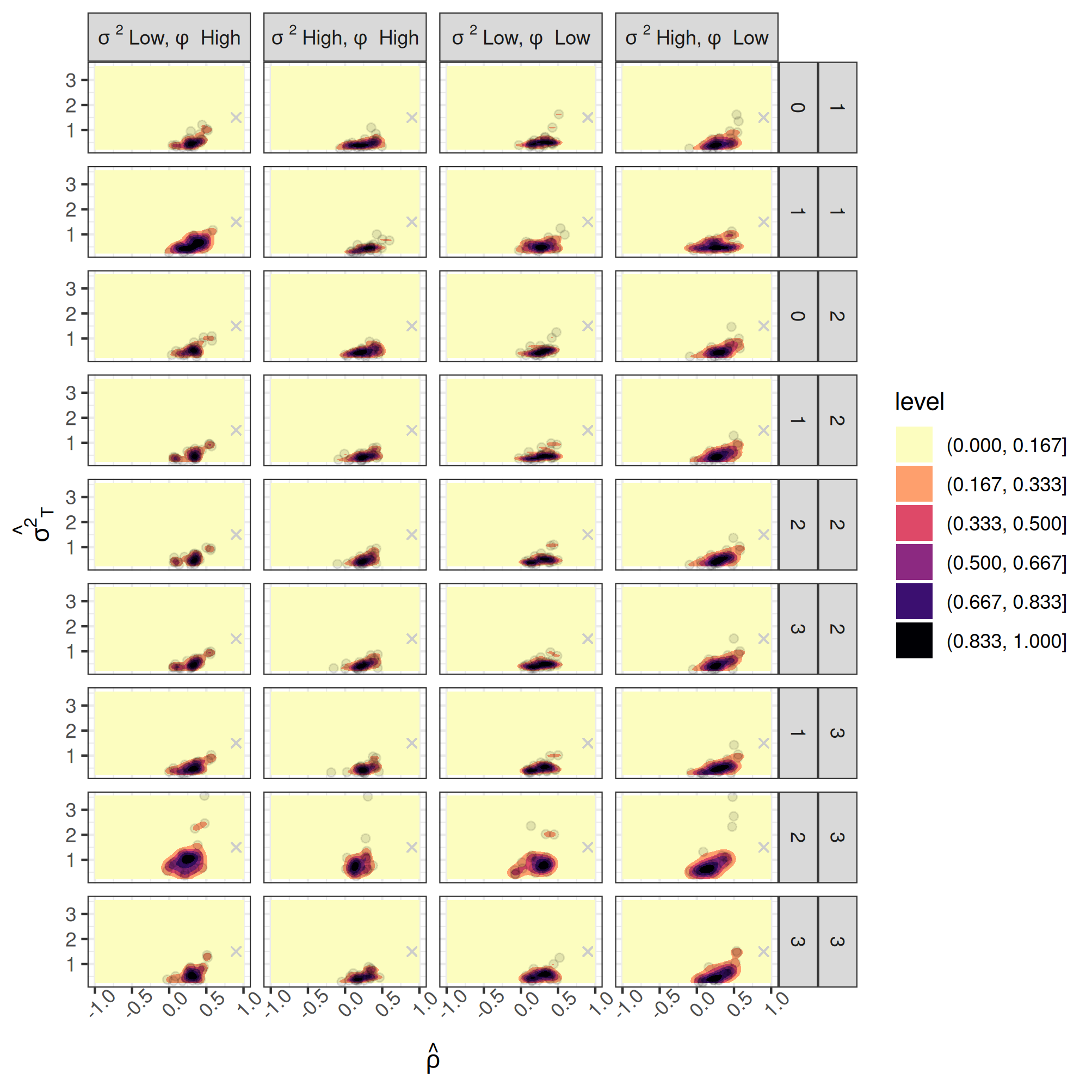}
    \caption{Contour plots showing the combination of point estimates of the temporal autocorrelation ($\hat{\rho}$, x-axis) and variance ($\hat{\sigma^2_T}$, y-axis). Studies and data design scenarios are shown across the facet rows, and spatial autocorrelation sub scenarios shown along facet columns. These results were produced by the sub scenarios with high temporal correlation $\rho=0.9$ and temporal variance $\sigma^2_T=1.5$ (gray crosses). The density is given by 100 simulation runs per dataset. `level' in the legend depicts the discretized proportion of runs within each contour plot region. The gray crosses depict the true parameter values.}   
    \label{fig:contour-rho-sigmasqT}
\end{figure}

Combinations of point estimates of the temporal autocorrelation coefficients $\hat{\rho}$ and $\hat{\sigma^2_T}$ were overall biased when temporal correlation and variance were both truly high (Fig. \ref{fig:contour-rho-sigmasqT}). Nonetheless, the bias was lower when levels of temporal correlation and variance were truly low (Fig. B.23). The MSE of $\rho$ estimator was overall constant across studies-scenarios 1-0 to 3-1, and showed an increase in the study-scenario 3-2 and 3-3 (Fig. B.24).

\subsection{Empirical data analysis}

\begin{figure}[htbp]
    \centering
    \includegraphics[width=1\linewidth]{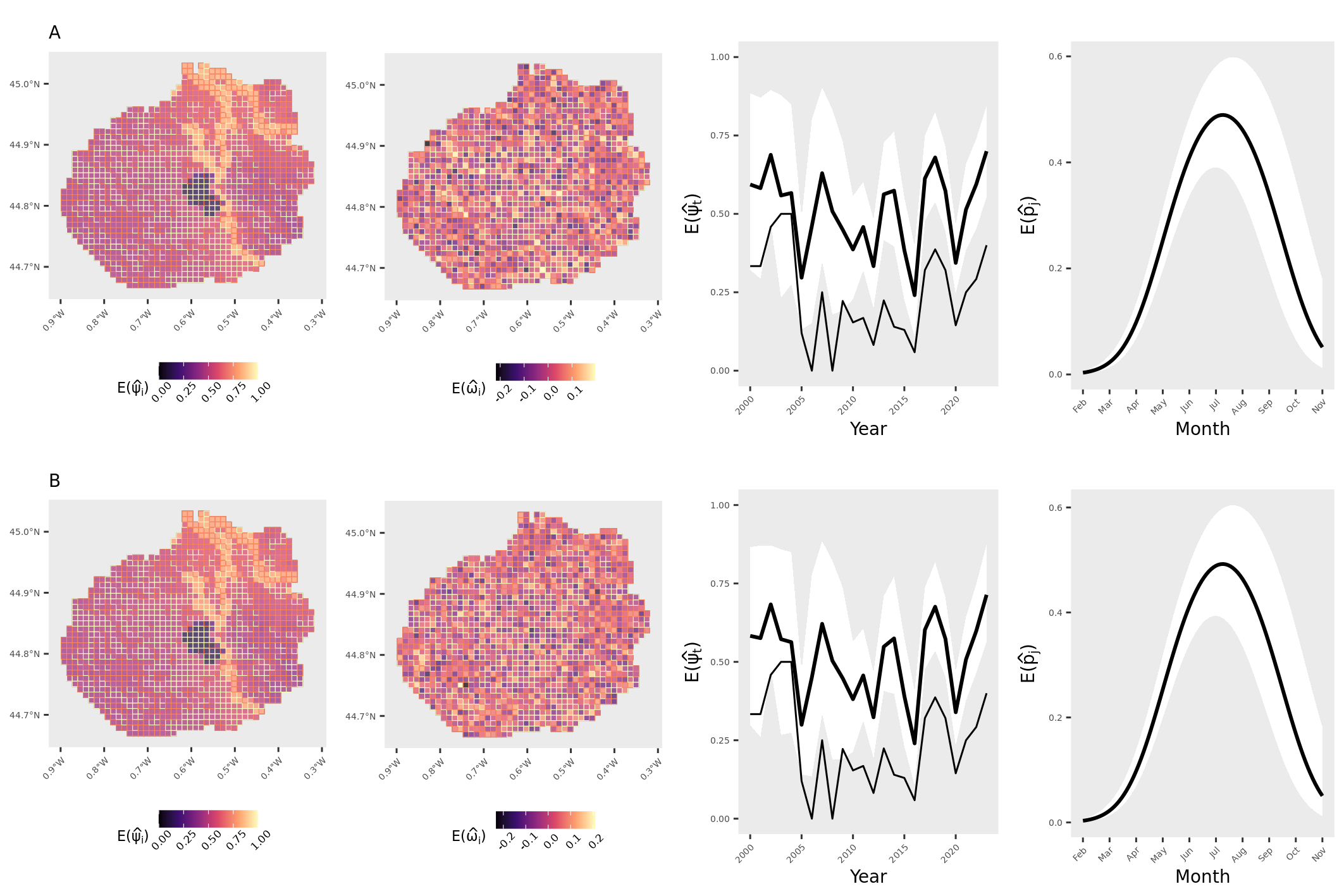}
    \caption{Distribution of the common blue \textit{Polyommatus icarus} in the greater Bordeaux area, Southwest France. (A) Results for the model with NNGP=15 and weakly informative prior for $\phi$. (B) Results for the model with NNGP=15 and an informative prior for $\phi$. Values plotted in the occupancy map (left) were calculated as $\mathbb{E}(\hat{\psi_{i}}) = {\frac{1}{T}\sum_{t=1}^{T}~\hat{\psi_{it}}}$, and refer to in-sample estimates for sites with data (cells with yellow borders) and out-of-sample predictions for missing cells (those with red borders). The spatial random effect $\mathbb{E}(\hat{\omega}_i)$ was also derived from in-sample estimates and out-of-sample predictions from the model. Trend plots show the proportion of occupied sites over the 24 years of data, calculated as $\mathbb{E}(\hat{\psi_{t}}) = {\frac{1}{I}\sum_{i=1}^{I}~\hat{\psi_{it}}}$. The naive yearly occupancy trend (number of sites with detection divided by the total number of sites sampled per year) is depicted by the thinner line. Detection probability over months (secondary periods) was also derived from in-sample predictions. The black line depicts the average detection trend, and the white bands depict the upper and lower bounds of the 95\% Credible Interval. No $1 \times 1$ km cell was 100\% covered by water (the maximum was 93\%). Parameter values derived from 1,200 posterior distribution draws.}
    \label{fig:p-icarus-results}
\end{figure}

Fitting the model to \textit{P. icarus} data showed that the average yearly site occupancy estimate (average of $\mathbb{E}(\hat{\psi_{t}})$) was 50.93\% (95\% Credible Interval: $[31.36\% - 71.24\%]$), against a naive average yearly occupancy of 24.12\% (average of 15.75 cells with detection per year). There were detections in a total of 227 cells across all the 24 years. We found a fluctuating occupancy trend of the common blue over time, which might be stable in the long term. This trend differed from the naive occupancy, which increased from 2010 to 2023, but decreased in the long term if we compare the first and last years (Fig. \ref{fig:p-icarus-results}A). The average detection probability per monthly survey was 46.03\% (95\% Credible Interval: $[37.70\% - 54.42\%]$). Occupancy decreased with elevation, urban and non-water cover. Latitude and longitude had a weak effect on common blue occupancy (Supporting Information C, Table C.1). Detection probability increased with the number of observers and with non-water land cover, and decreased with latitude. Detection peaked during aural summer months (Fig. \ref{fig:p-icarus-results}A).

Regarding prior-posterior overlap (PPO), the results showed that the intercept, coefficients of latitude, non-water cover, temporal variance $\hat{\sigma^2_T}$ and correlation $\hat{\rho}$ had high PPO (close or above 30\%, Table C.1). The estimate of $\hat{\phi}$ was large $\hat{\phi}=32.49 ~[4.56 - 58.70]$, indicating very short autocorrelation range \(\frac{3}{\hat\phi}\). The low spatial autocorrelation is depicted by the map of spatial random effects, which shows no recognizable spatial pattern (Fig. \ref{fig:p-icarus-results}A).

\begin{figure}[htbp]
    \centering
    \includegraphics[width=1\linewidth]{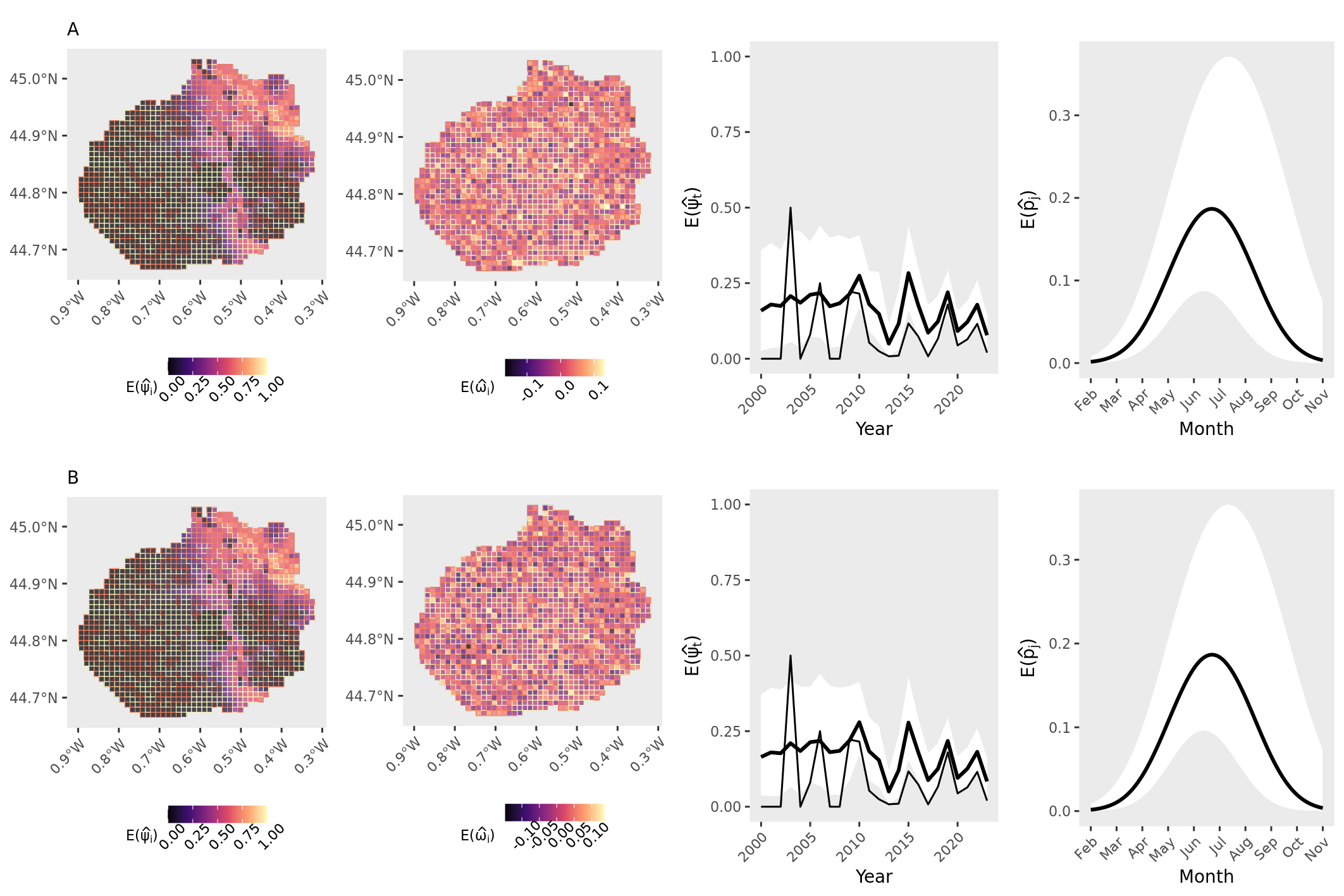}
    \caption{Distribution of the large copper \textit{Lycaena dispar} in the greater Bordeaux area, Southwest France. (A) Results for the model with NNGP=15 and weakly informative prior for $\phi$. (B) Results for the model with NNGP=15 and an informative prior for $\phi$. Values plotted in the occupancy map (left) were calculated as $\mathbb{E}(\hat{\psi_{i}}) = {\frac{1}{T}\sum_{t=1}^{T}~\hat{\psi_{it}}}$, and refer to in-sample estimates for sites with data (cells with yellow borders) and out-of-sample predictions for missing cells (those with red borders). The spatial random effect $\mathbb{E}(\hat{\omega}_i)$ was also derived from in-sample estimates and out-of-sample predictions from the model. Trend plots show the proportion of occupied sites over the 24 years of data, calculated as $\mathbb{E}(\hat{\psi_{t}}) = {\frac{1}{I}\sum_{i=1}^{I}~\hat{\psi_{it}}}$. The naive yearly occupancy trend (number of sites with detection divided by the total number of sites per year) is depicted by the thinner line. Detection probability over months (secondary periods) was also derived from in-sample predictions. The black line depicts the average detection trend, and the white bands depict the upper and lower bounds of the 95\% Credible Interval. No $1 \times 1$ km cell was 100\% covered by water (the maximum was 93\%). Parameter values derived from 1,200 posterior distribution draws.}
    \label{fig:l-dispar-results}
\end{figure}

Fitting the model to \textit{L. dispar} data showed that the averaged yearly site occupancy estimate was 12.85\% (95 \% Credible Interval: $[2.72\% - 34.54\%]$), against a naive average yearly occupancy of 8.54\%(average of 4.54 cells with detection per year). There were detections in a total of 70 cells across all the 24 years. We observed an uncertain occupancy trend of the large copper before 2010. There was a fluctuating occupancy trend during the subsequent years, although maximum occupancy has decreased in recent years. The naive yearly occupancy trend fluctuated greatly and was below the estimated yearly occupancy in most of the years (Fig. \ref{fig:l-dispar-results}A). The average detection probability per monthly survey was 18.5\% (95\% Credible Interval: $[8.67\% - 34.6\%]$). Occupancy decreased with elevation, and increased with latitude, longitude, and marsh cover (Table C.2) despite the decline at high marsh cover levels. Detection probability increased with the number of observers, and decreased with latitude and non-water land cover (Table C.2). Detection peaked during aural summer months (Fig. \ref{fig:l-dispar-results}A).

Regarding PPO, the results showed that the estimates of coefficients of latitude, longitude, marsh cover, and urban effect, as well as spatial variance $\hat{\sigma^2}$, temporal variance $\hat{\sigma^2_T}$ and correlation $\hat{\rho}$ had high PPO, close or above 30\% (Table C.2). The longitude and urban-cover coefficient, as well as the spatial and temporal variance parameters, did not converge across chains (Table C.2). As for the common blue, the estimate of $\hat{\phi}$ was large $\hat{\phi}=31.63~[4.41 - 57.93]$, depicting a short spatial autocorrelation range (Fig. \ref{fig:l-dispar-results}A). 

\subsection{Sensitivity analyses}

Fitting the model with spatially uncorrelated site random effects to truly spatially autocorrelated datasets (study 2-scenario 1) yielded a more biased $\psi_{it}$ estimator than models accounting for spatial autocorrelation. Overall, the best performance of this model occurred when spatial variance was low (Fig. F.1). In the low spatial autocorrelation scenario (high decay $\phi$), the model could not capture truly existing patches of occupancy (Fig. F.2). Also the model overestimated occupancy when it was truly low, and underestimated otherwise (Fig. F.1). The intercepts and regression slopes of occupancy and detection models were not estimated with bias (Fig. F.3). These results are shown in our GitHub page (Supporting Information F, see Data Availability Statement).

Fitting occupancy models with NNGP=15 neighbors to data of study-scenario 3-2 (Fig. E.14) and 3-3 (Fig. E.15), and using a larger observation spot ($I^*=600$ sites) (Fig. E.16), produced results similar to those shown in Fig. \ref{fig:study3-scenario3-results}.

Regarding the sensitivity analysis applied to empirical data collected within the greater Bordeaux area, the parameters of occupancy and detection models ($
\boldsymbol{\beta}$ and $\boldsymbol{\alpha}$) changed only subtly for both species when using an informative prior for $\phi$ (Figs. \ref{fig:p-icarus-results}B and \ref{fig:l-dispar-results}B; Tables C.1 and C.2). The spatial decay $\hat{\phi}$ was strongly constrained by the informative prior, showing a PPO higher than 93\% (Tables C.1 and C.2). Notably, the spatial random effects $\hat{\omega_i}$ resembled each other across analyses with either weak or informative priors. In all cases, spatial random effects indicated short autocorrelation range (Fig. \ref{fig:p-icarus-results}A-B; Fig. \ref{fig:l-dispar-results}A-B; Figs. C.1-C.2). Similar results were obtained in the analyses of data of the four remaining species (Supporting Information G).

The analysis of the full Nouvelle-Aquitaine dataset resulted in similar issues regarding the estimation of spatial and temporal parameters (Figs. D.1-D.4, Tables D.1-D.2). For the common blue, random effect estimates indicated short-range autocorrelation among missing cells (Fig. D.1). When mapping the estimated spatial random effects for both missing and non-missing cells, a flat pattern emerged with spatial random effect values mostly constant (close to zero) in space, which resulted in high estimated occupancy ($\mathbb{E}(\hat{\psi_i}) \geq 0.5$) across most of Nouvelle-Aquitaine (Fig. D.3). Model predictions for the common blue indicated low occupancy probability in the North (Poitiers) and areas of higher elevation (Pyrenees in the South, Limousin in the Northeast), and high occupancy elsewhere (Figs. D.1 and D.3). There was a declining trend of common blue occupancy over time; the naive yearly occupancy was below the estimated occupancy, and showed a similar declining trend (Fig. D.3). A detection peak was found in mid-July, and the estimates were more precise than when using data from the greater Bordeaux area (Fig. \ref{fig:p-icarus-results}). 

Occupancy of the large copper was low overall, being $\mathbb{E}(\hat{\psi_i}) \leq 0.3$ across most of Nouvelle-Aquitaine. Occupancy was high ($\mathbb{E}(\hat{\psi_i}) \geq 0.75$) only in sites where the species was detected (Fig. D.4). This pattern was caused by near zero random effect estimates when considering the full dataset (Figs. D.2 and D.4). Overall, low occupancy of the large copper occupancy was found in Landes (west of Nouvelle-Aquitaine), along the Pyrenees, and in Limousin (Fig. D.4). Higher occupancy was estimated along rivers -- wetlands of the Adour river (Atlantic Pyrenees/western Pyrenees, south of Nouvelle-Aquitaine), Garonne and Dordogne rivers (around Bordeaux, center of Nouvelle-Aquitaine), and the Vienne river (Poitiers, north of Nouvelle-Aquitaine). But these patterns mostly reflected the observed occupancy data. The intercept and three regression slopes, as well spatial and temporal variance parameters, did not converge for this species (Table D.2). Yearly occupancy was low and showed a decreasing trend over time (Fig. D.4). Detection probability peaked in mid-June (Fig. D.4). Across all analyses, we found no serious problem for parameters of the detection model (Tables C.1-2, D.1-2). 

\section{Discussion}

We assessed the identifiability and estimation quality of a Bayesian multi-season occupancy model with spatial and temporal random effects \citep{doser2024fractional}, developed to alleviate challenges due to the absence of comprehensive spatial and temporal replication in naturalist observation databases. Using three empirically-motivated simulation studies and one empirical data analysis of the occupancy of two butterfly species, we evaluated the effects of (1) a skewed distribution of survey numbers per grid cell, including missing data (0 surveys), (2) overlap in detection and occupancy covariates, and (3) clustered observations in space and/or time. In addition, we evaluated model robustness to a mild misspecification case (probit to simulate data, logit to fit the model to data). 

With a quantity of data exactly equal to \citet{doser2024fractional} -- i.e., same average number of surveys -- we demonstrated that neither a skewed distribution of survey numbers nor an overlap of covariates (between occupancy and detection models) led to poorer estimation, compared to previously used one-or-two and/or one-or-four secondary occasions designs (single survey + $x$ replication within primary occasions, \citet{doser2024fractional, vonhirschheydt2023mixed}, respectively). Furthermore, robustness was maintained when the model was misspecified. Unlike what we originally thought when designing the simulation studies, under a heterogeneous distribution of surveys to sites, the model with autocorrelated random effect performed well in differentiating the effect of the same covariate on occupancy and detection. This is good news for those interested in fitting  D\&S' model to their own data. 

While the overlap of covariates between the occupancy and detection models was already shown to represent a challenge for the identifiability of occupancy models \citep{lele2012dealing}, our results show that this overlap ($L_i$ in both occupancy and detection models) did not cause bias on $\psi_{it}$ and regression coefficients, relative to situations of no to partial overlap of covariates. In other words, the inclusion of latitude in both occupancy and detection models was not enough to deteriorate the quality of the site-occupancy estimation. Thus, the inclusion of temporal and spatial autocorrelation in the model, and the replication level contained in the heterogeneous simulated data, despite being skewed to zero or one visit, were conditions that enabled the estimation of the model.

Other studies, applying different models, also have shown promising results for occupancy estimation when there is overlap of covariates between occupancy and detection submodels. For instance, \cite{hepler2018identifying} find that dynamic occupancy models (with autocorrelation in space and time), either with or without overlap of covariates, performed similarly when the number of primary occasions was large ($T=10, 20$, and $30$), both with data that had one single survey occasion and with more survey occasions ($J=3$ and $J=5$ per primary occasion). Using dynamic temporal autologistic occupancy models, which enabled the simultaneous estimation of occupancy, detection, colonization and extinction probabilities, \cite{peach2017single} find that with single-survey (atlas-like) data it is possible to recover all probabilities without bias when there was total and partial overlap of covariates, given that detection probability was modeled with a power term of effort. The models of \cite{doser2024fractional} and ours, and of \cite{peach2017single} and \cite{hepler2018identifying}, are different in terms of autocorrelation structure and sampling design used to simulated data. For \cite{doser2024fractional} and our study, spatially and temporally independent random effects were taken from Gaussian Distributions, and simulated data was heterogeneous in the sense that levels of survey replication varied across sites. In \cite{peach2017single} and \cite{hepler2018identifying}, an autologistic autocorrelation model was used, which means that occupancy in $t$ depends on the spatial neighborhoods' occupancy at $t-1$, and survey replication was constant across sites. Taken together, these studies demonstrate that the overlap of covariates between the occupancy and detection submodels is a minor problem for occupancy estimation when autocorrelation is incorporated into the model, either for heterogeneous or homogeneous replication levels. The ability to differentiate the effects of covariates on occupancy and detection is one of the strengths of occupancy models \citep{lahoz-monfort2014imperfect}, yet it is rarely used in practice \citep{goldstein2024ecologists}.

Despite these encouraging findings, we found identifiability issues elsewhere in the model. In spatial models as the one used here, the spatial decay parameter $\phi$ (which controls the autocorrelation range) and the spatial variance parameter $\sigma^2$ (which controls the magnitude of the spatial variability) are theoretically weakly identifiable \citep{zhang2004inconsistent, Doser2023Convergence}. It means that it is not always possible to uniquely recover the data generating process, since several $\phi$ and $\sigma^2$ values can yield data with the same likelihood \citep{cole20202parameter}. This behavior is exemplified in our density plots. In scenarios of high $\phi$ and low variance $\sigma^2$, the density of $\hat{\phi}$ and $\hat{\sigma^2}$ combinations had an elongated (flat) shape, and sometimes two density spots occurred along the range of $\phi$ values. Multiple high-density spots of estimates were also evidenced for scenarios with high $\phi$ and $\sigma^2$ and low $\phi$ and high $\sigma^2$.

Issues regarding the identifiability of spatial models are not new. For instance, issues with spatial random effects and occupancy predictions were found by \citet{latimer2006building} in an exponential autocorrelation model similar to the one used here. Also, \citet{datta2016hierarchical} showed that for sparse data---where a cell/site lacks neighbors and sampled sites are distant from each other---the nearest neighbor Gaussian Process (NNGP) covariance function cannot efficiently represent the covariance function of a full Gaussian process. Under this condition, \citet{datta2016hierarchical} found low autocorrelation estimates (high $\hat\phi$) and out-of-sample predictions that just reflected this limited sharing of information between sites. Considerations about the weak identifiability of spatial autocorrelation parameters in \texttt{spOccupancy} models were also made by \citet{Doser2023Convergence}, which advises using informative priors to minimize identifiability problems. In addition, the overestimation of the spatial decay $\phi$ is in accordance with the findings of \citet{belmont2024spatiotemporal}. They suggest that the NNGP approach and the use of sparse matrices \citep{datta2016hierarchical} is too spatially restrictive to account for spatial dependence at large distances, and showed that a multi-season occupancy model with a full Gaussian field (implemented in R-INLA) can efficiently recover $\phi$ under a strong autocorrelation situation. In another assessment, \citet{zhang2004inconsistent} found identifiability issues in a Matérn-class spatial model, where a flat likelihood of the spatial correlation parameter $\theta$ was found when the spatial variance $\sigma^2$ was enabled to be estimated by the model. However, $\theta$ was identifiable when $\sigma^2$ was fixed. Furthermore, it was found that the ratio $\frac{\theta}{\sigma^2}$ was identifiable and could be a useful model parametrization when the study goal is interpolation. Finally, a more recent study found bias in spatial decay and variance estimation in a spatial model using Gaussian process with Matérn covariance function \citep{makinen2022spatial}. These findings show that there are often fundamental identifiability issues in the formulation of spatial models, in the sense that autocorrelation and spatial variance parameters may not be individually estimable. 

Weak identifiability does not necessarily imply bias on spatial and temporal autocorrelation parameters, but we did find some as well. The spatial decay estimator was biased high when it should be low, so that the random effects appeared invariably to have little or no spatial autocorrelation (like unstructured random effects), situations that cannot be differentiated by the model \citep{Doser2023Convergence}. Thus, correlation in occupancy probability abruptly dropped with the geographic distance between sites. Furthermore, the $\phi$ estimator was also imprecise, with $\hat{\phi}$ values ranging from 4 to almost 30, thus covering half of the prior-distribution range $U(3,60)$ and indicating that the model has difficulties to update prior information with the data. Another concerning result was the biased estimation of temporal autocorrelation and variance $\rho$ and $\sigma^2_T$. Their estimation was biased low across all autocorrelation scenarios. Thus, temporal random effects might look unstructured and result in unreliable estimates of annual site occupancy, as they may show more temporal variation in occupancy (or less similarity in occupancy between adjacent years) than is actually the case \citep{outhwaite2018prior}. In sum, these results indicate that this multi-season occupancy model with spatial and temporal autocorrelation tends to indicate little to no spatial and temporal autocorrelation when they truly exist.

What are then the consequences of weak identifiability and bias in the spatiotemporal random effects model for estimated occupancy? The consequences were well visualized when predictions were needed in the last simulation scenarios where simulated occupancy data was clustered in space and time. This clustered configuration of surveys is a common characteristic of the datasets used to model species distributions \citep{altwegg2019occupancy, bowler2024treating}, and can be generated when, for instance, fieldwork takes place in locations closer to where most people live and/or in the vicinity of attractive locations \citep{isaac2020data}, and occurs during periods when the focal species is more likely to be sighted \citep{bishop2013utility}. Such spatially and temporally clustered observations yielded spatial random effects with small variation, and a biased high occupancy estimator. There was an overestimation of the occupancy estimator when it was truly low, and the site-level occupancy estimates were closer to the yearly average site occupancy ($\beta_0$) than they should have been. This pull towards the average occurred due to the strong influence of the random effects, combined with a biased high spatial decay estimation and a biased low temporal autocorrelation. Predictions of occupancy in space for unsampled sites were thus nearly constant, reflecting the average of the spatial random effect. Interestingly, a similar pattern was found in the analyses of the common blue occupancy based on the full Nouvelle-Aquitaine data. D\&S (p. 366) suggested that ``future simulation studies could assess the reliability of `mixed' designs when there is a non-random spatial and/temporal pattern in the sites and/or seasons in which multiple visits are performed''. We provide here this assessment, and find that spatiotemporal clusters in occupancy data are challenging for occupancy models with spatially and temporally autocorrelated random effects.

We explored some `solutions' to the identifiability issues. The first one was to simply get rid of autocorrelation parameters by considering an alternative model with i.i.d. random effects, in order to evaluate if a simpler model would perform best. This model was initially developed to estimate regional-level and country-level temporal occupancy trends using large and sparse data \citep{outhwaite2018prior}. Fitting this model to truly spatially and temporally autocorrelated simulated data, under a relatively benign setup---no overlap of covariates between the occupancy and detection models---did not show promising results, especially when spatial autocorrelation and variance were high. The second was to use more neighbors in the Gaussian Process approximation, which did not solve the issue. When analyzing the empirical data, we also tried an informative prior for $\phi$ in the spatially autocorrelated model \citep{Doser2023Convergence, bajcz2024within}, which did not improve parameter estimation. We fitted models to the full Nouvelle-Aquitaine empirical dataset in addition to the subset that worked best, but the identifiability issues were still there in the full dataset, and spatial autocorrelation was estimated as non-existent. We could assume, of course, that the empirical dataset (unlike our simulations) is spatially uncorrelated and then get rid of the spatial random effects. However, doing so would go against the knowledge on butterfly metapopulations accumulated so far, and the occupancy maps resulting from this omission of spatial autocorrelation would not be reliable. As an alternative statistical framework, CAR models were tested in preliminary analysis \citep{latimer2006building, hepler2021spatiotemporal}, but it was computationally prohibitive to build a spatial neighborhood for 90,000+ sites. Other alternatives (not tested here) include the recently developed INLA models that use the full Gaussian random field to generate the random effects \citep{hepler2021spatiotemporal, belmont2024spatiotemporal}, and other model parameterizations \citep{zhang2004inconsistent} that would require a full model rethink. 

Recent developments in occupancy modeling intend to deliver computationally efficient models using spatial and temporal autocorrelation. Their use is justified by the need to alleviate the lack of replication in occupancy data while enhancing model predictive performance \citep{johnson2013spatial, hepler2018identifying, altwegg2019occupancy, diana2023fast, belmont2024spatiotemporal, dennis2024efficient, doser2024fractional}, building on the fact that adjacent sites and years share information about occupancy and/or detection \citep{johnson2013spatial}. While the approach sounds promising, and may well become routine in future years, we found that in the current models such as the one of \citet{doser2024fractional}, spatiotemporally correlated random effects combined with spatiotemporal imbalance in the distribution of records/effort substantially impact model performance. In our empirical example, for this reason it was not possible to obtain reliable parameter estimates over the whole study area (shown in Supporting Information). A focus on well-studied data subset (shown in main text) was more promising, in the sense that it produced sensible average site occupancy and annual occupancy estimates, but was apparently still prone to identifiability issues for spatiotemporal autocorrelation parameters. Thus, we conclude at the present time that while occupancy models with spatiotemporal autocorrelation are robust to a heterogeneous sampling effort and covariate overlap between submodels, they are prone to practical identifiability issues and only applicable in the absence of severe data gaps in space and time, whose presence tends to contaminate predictions even in data-rich areas. 

\subsection*{Conflict of interest statement}
We have no conflicts of interest to disclose.

\section*{Acknowledgments}
This work was supported by the cooperation OFB-22-1513 between INRAE (French Agricultural and Environmental Institute) and OFB (French Office of Biodiversity). FB acknowledges support from Bordeaux Métropole. All authors acknowledge the support from the Nouvelle-Aquitaine Wildlife Observatory (FAUNA). We thank Fr\'ed\'eric Archaux and Fabien Laroche for detailed comments on the manuscript, as well as Fr\'ed\'eric Gosselin and Lise Maciejewski for contributing during discussions. A list of contributors to the whole dataset can be found in the FAUNA website: \url{https://observatoire-fauna.fr/programmes/portails-taxonomiques/papillons-de-jour}. The raw names/IDs of contributors of data used in the present study can be found in our GitHub page - Online Supporting Information H. We warmly thank all the individuals and institutions who collected data. 

\section*{Data Availability Statement}
A version of the dataset, all codes used in simulations and empirical data analyses, and the supporting information are available on GitHub (\url{https://github.com/andreluza/butterfly_occupancy.git}). The dataset, supporting information, all codes used in simulations and empirical data analyses, and the results (RData) are also available on Zenodo (\url{https://doi.org/10.5281/zenodo.19633172}). Habitat covariates (CORINE) and elevation (EU-DEM) were downloaded from \url{https://inpn.mnhn.fr/habitat/cd_typo/22} and \url{https://sdi.eea.europa.eu/catalogue/srv/api/records/3473589f-0854-4601-919e-2e7dd172ff50}, respectively. The $1 \times 1$ km spatial grid was downloaded from: \url{https://observatoire-fauna.fr/ressources/publications?typePublication%5B%5D=fauna&themesID%5B%5D=6}. 

\bibliography{JABES/1-Manuscript}
\bibliographystyle{JABES/ecol_let}

\end{document}